\documentclass[11pt,eqs]{article}
\usepackage{latexsym}

\def\cleq{\setcounter{equation}{0}}

\makeatletter
\newcommand\xleftrightarrow[2][]{%
  \ext@arrow 9999{\longleftrightarrowfill@}{#1}{#2}}
\newcommand\longleftrightarrowfill@{%
  \arrowfill@\leftarrow\relbar\rightarrow}
\makeatother

\title{From geometry to non-geometry via T-duality
\thanks{Work supported in part by
the Serbian Ministry of Education and Science, under contract No. 171031.}}
\author{B. Sazdovi\'c
\thanks{e-mail: sazdovic@ipb.ac.rs}\\
{\it Institute of Physics,}\\
{\it University of Belgrade,}\\
{\it 11001 Belgrade, P.O.Box 57, Serbia}}

\begin{document}
\maketitle
\begin{abstract}
Reconsideration of the T-duality of the open string allows us to introduce some geometric features in non-geometric theories.
First, we have found what symmetry is T-dual to the local gauge transformations.
It  includes transformations of background fields but does not include  transformations of the coordinates.  According to this  we have introduced  a new, up to now missing term, with additional gauge field $A^D_i$ (D denotes  components with Dirichlet boundary conditions). It compensates non-fulfilment of the invariance under such transformations on the end-points of an open string, and the standard gauge field $A^N_a$ (N denotes components with Neumann  boundary conditions) compensates non-fulfilment of the gauge invariance. Using a generalized  procedure we will  perform T-duality of  vector fields  linear in coordinates. We show that gauge fields $A^N_a$ and $A^D_i$ are T-dual  to ${}^\star A_D^a$ and  ${}^\star A_N^i$ respectively.

We  introduce  the field strength of T-dual non-geometric theories as derivatives of T-dual gauge fields along both T-dual variable $y_\mu$ and its double ${\tilde y}_\mu$.
This definition  allows us to  obtain  gauge transformation of non-geometric theories which leaves the T-dual field strength invariant. Therefore, we
introduce some new features of non-geometric theories where field strength has both antisymmetric and symmetric parts. This allows us to define new kinds of truly non-geometric theories.
\end{abstract}

%%%%%%%%%%%%%%%%%%%%%%%%%%%%%%%%%%%%%%%%%%%%%%%%%%%%%%%%%%%%%%%%%%%%%%%%%%%%%%%%%%%%%
\cleq

\section{Introduction}

String theory has  more symmetries  than point particle theory. This is the  source of an unusual situation, which is described by so-called non-geometry \cite{HMW,DH,STW}. In fact, when going around a loop
in space-time the field configuration is well defined only after   applying some  string symmetry (T-duality) as a transition function.

Geometric spaces  appear when diffeomorphisms and gauge transformations have been used as transition functions to   overlap  coordinate patches. According to Ref.~\cite{STW}, there are two kinds of non-geometric backgrounds. In the first case (a benign form, which for a three-torus is usually referred to as a theory with Q-flux) the background is locally  geometric but globally non-geometric. This is
 T-fold , when  T-duality transformations can be used as transition functions \cite{H}. In the second case
(a severe form, which for a three-torus is usually referred to as a theory with R-flux)  we lose the local geometric description of space-time points and the background is  non-geometric even locally.
This is a mysterious background, when T-duality is performed along  some  non-isometry directions.

In the great majority of papers, Abelian T-duality has been applied  along the coordinates with global shift symmetry. A problem occurs when  we  try  to perform T-duality along the coordinates
on which background fields depend. Then we should  apply the generalized Buscher's procedure, developed
in Refs.~\cite{DS1,DS2}, where the metric and Kalb-Ramond fields are   coordinate-dependent. In that case, the argument of T-dual background fields
is not simply the T-dual variable $y_a$ but it is the line integral of world-sheet gauge fields $v_{+}^a$ and  $v_{-}^a$. Explicitly, we have
$V^a [v_{+},v_{-}]\equiv \int_{P}d\xi^\alpha v^{a}_{\alpha} =\int_{P}(d\xi^{+} v^{a}_{+} +d\xi^{-} v^{a}_{-}) $, which on the solution for gauge fields turns to
$V^a = - \kappa \, \theta^{a b } y_b + G^{-1 a b}_E \, {\tilde y}_b$, where $\dot{\tilde y}_a = y^ \prime_a$ and ${\tilde y}^\prime_a = \dot y_a$.
We will  claim   that in such cases  T-dual theories are locally non-geometric, although the initial theory is geometric.
So far we have two reasons for that.  Besides the fact that we perform T-dualization  along non-isometry directions,
we obtain a locally non-geometric background (the argument is the line integral).
The additional two features: non-commutativity  of the closed string coordinates and non-associativity, which have been shown in
Refs.\cite{DNS,DNS2}.

In order to better understand non-commutativity, we prefer to use a canonical method. It can help us to  reproduce nicely the main result of open string theory and non-commutative geometry,
discovered in Ref.~\cite{SW}. In fact, we can solve the Neumann boundary condition and obtain an effective theory with effective canonical variables $q^i$ and $p^i$ and effective background, named the open string background in
Ref.~\cite{SW}. Then, the initial coordinate $x^i$ depends not only on effective coordinates but also on effective momenta,  $x^i = q^i -2 \theta^{i j} \int^\sigma d \sigma_1 p_j $. Since $q^i$ and $p_i$, as independent
variables, satisfy standard Poisson brackets, it is clear that initial coordinates non-commute.
In Ref.~\cite{SA1} it is described  how an antisymmetric field $B_{a b}$, regarded as a magnetic field on the Dp-brane, has  an effect on non-commutativity.

In Refs.~\cite{DL,BDLPR} it has been shown that the commutators of  closed string coordinates  are proportional to the flux and winding number. For the relation between space-time symmetry and non-commutativity see Ref.~\cite{SA2}  and references therein.  We can expect that in the canonical formulation closed string coordinates, in order to be  non-commutative, should have similar expressions
to the open string ones. However, closed string coordinates do not have  end points and so there are no boundary conditions and so no effective coordinates.

In Ref.~\cite{DNS}  this problem has been  solved by T-duality performed along non-isometry directions, so that T-dual background fields depend on $V^\mu$.
To understand the essence it is useful to perform the Buscher
procedure on the Lagrangian in canonical form  $S= \int d^2 \xi [\pi_\mu {\dot x}^\mu - {\cal H}(x, x^\prime, \pi) ]$. The $x$-dependence comes from the arguments of background fields.
Then the corresponding auxiliary action (after gauging shift symmetry, putting the corresponding
field strength to zero and fixing the gauge, see Section~3.3.1) takes the form $S_{aux} = \int d^2 \xi [\pi_\mu v_0^\mu - {\cal H}(V, v_1, \pi) - \kappa(v_0^\mu y^\prime_\mu - v_1^\mu {\dot y}_\mu)]$.
Formally, we substitute ${\dot x}^\mu \to v_0^\mu$, $x^\mu \to V^\mu$ and add the last term.
 Now, we can consider two cases. First, when we work only with isometry directions then we have ${\cal H}(v_1, \pi)$ (the Hamiltonian ${\cal H}$ does not depend on $x^\mu$ and consequently on $V^\mu$).
Varying with respect to $v_0^\mu$ we obtain the  relation $ \kappa y^\prime_\mu = \pi_\mu$. (After integration over $\sigma$ it produces a well known relation
between momenta and winding numbers). This relation cannot help us, because T-dual variables $y_\mu$ commute, as they  depend only on initial momenta and not on initial coordinates.
The second case, when we work  with non-isometry directions, is more interesting.
Then the argument of T-dual background fields $V^\mu$ is  a non-local expression because  it is a line integral of both
$v_0^\mu$ and $v_1^\mu$.
 So, variation with respect to $v_0^\mu$ produces a new term and we  obtain  $\kappa y_\mu^\prime = \pi_\mu - \frac{\kappa}{3} B_{\mu \nu \rho}x^{\prime \nu} x^\rho $ (see relation (2.22) of Ref.~\cite{DNS}).
Now, the T-dual variables $y_\mu$ depend on both initial variable $x^\mu$ and its canonically conjugated momentum $\pi_\mu$, which provides non-commutativity between T-dual variables $y_\mu$. The fact that
this expression is quadratic in coordinates provides non-associativity  between T-dual variables $y_\mu$, see Ref.\cite{DNS2}.

In the present article we are going to perform T-dualization  for a background with constant metric and Kalb-Ramond fields but where
the vector gauge fields are linear in coordinates. We will do this in two ways: in terms of vector  fields and in terms of corresponding field strengths. In the first case,  T-dual gauge fields will depend on the same expression $V^\mu$, introduced above. So in this case, due to the presence of vector background fields,  the T-dual theory will be non-local and hence locally  non-geometric.
On the other hand, this theory can be described as a theory with constant field strength, but with both antisymmetric and symmetric parts of field strength.
So, in the second case, we can perform standard Buscher T-duality and obtain explicitly T-dual field strengths. The main contribution
of the paper is the relation between T-dual gauge fields and T-dual field strengths. It is non-standard for two reasons. First,   because we must use derivatives of vector fields with respect to two variables: the T-dual variable $y_\mu$ and its double ${\tilde y}_\mu$. Second, because the T-dual field strength contains both antisymmetric and symmetric parts.
So, using T-duality we are able to introduce some geometry (the field strength in terms of gauge fields) for non-geometric theories.

To prepare this, we will first reconsider the T-duality of vector background fields.
There is a standard way to introduce vector fields at the end of an open string (see for example Ref.~\cite{Zw}). In fact, gauge invariance of the Kalb-Ramond field $B_{\mu \nu}$,
which is valid for  the closed string, has failed on open string ends with Neumann boundary conditions. To restore it we should add the corresponding vector fields $A_a$ at the string end-points.
Then neither the Kalb-Ramond field $B_{a b}$ nor the field strength $F_{a b}= \partial_a A_b - \partial_b A_a$ are  gauge invariant. There is a new invariant quantity ${\cal B}_{a b}= B_{a b}+ F_{a b}$.

We will show that there exists a procedure T-dual  to that explained in the previous paragraph. The main point is to understand what is T-dual to the local gauge invariance of the Kalb-Ramond field $B_{\mu \nu}$. In the space-time
formulation it is known that this is general coordinate transformations, but the world-sheet action is invariant under general coordinate transformations. It does not fail on open string ends with Dirichlet
boundary conditions, as we need. We will show that a transformation which includes transformation of background fields but not transformation of coordinates, is really  T-dual to the local gauge invariance. The closed string is invariant under such symmetry on the equations of motion. In this article we are going to show that this symmetry   fails on open string ends with Dirichlet boundary conditions.
In analogy with previous case we will   introduce a corresponding additional vector field ${A}_i$, which restores general coordinate transformations   at the string end-points.

In accordance with the boundary conditions, we will rename the vector fields $A_a$ to $A_a^N$ and  ${A}_i$ to  $A_i^D$,  where $A_a^N$
are fields corresponding to the Neumann boundary conditions and  $ A_i^D$  are fields corresponding to the Dirichlet  boundary conditions.

The Dirichlet vector field is not coupled with $\dot x^\mu$ but with expression $\gamma^{(0)}_\mu (x)$, which depends on both $\dot x^\mu$ and $x^{\prime \mu}$.
We will introduce $\gamma^{(0)}_\mu (x)$ in Section 2 and we will call it $\sigma$-momentum, because standard momentum  $\pi_\mu$  and $\gamma^{(0)}_\mu (x)$ are components of the same world-sheet vector.  Consequently, we introduce a pair of {\it effective vector fields} ${\cal A}_{\alpha \mu} \,\, (\alpha=0,1)$   as a world-sheet and space-time vector. Its world-sheet components: the standard one ${\cal A}_{0 \mu}$ is a coefficient in front of $\dot x^\mu$ and a new one ${\cal A}_{1 \mu}$ is a coefficient in front of $x^{\prime \mu}$. We will show that the  field strength
corresponding to ${\cal A}_{0 \mu}$ is antisymmetric while the non-standard one, corresponding to ${\cal A}_{1 \mu}$, is symmetric.

The space-time  equations of motion in the lowest order in slope parameter $\alpha^\prime$ are a consequence of the requirement of world-sheet conformal invariance on the quantum level.
We will consider the simplest  solutions for the closed string background fields (metric and Kalb-Ramond field) which satisfy the space-time equations of motion
$G_{\mu\nu}=const, B_{\mu\nu}  = const$. For gauge fields we will choose non-trivial solutions of the  space-time equations of motion \cite{Le}: we will take them  linear in coordinates with infinitesimal coefficients,
so that the field strength is infinitesimal and  constant. This is a non-trivial generalization of the standard consideration in the literature. As is well known \cite{Pol}, the constant part of the Dirichlet vector field
$A^D_i$ carries out uniform translation of the Dp-brane. In the present article the vector field additionally contains an infinitesimal coordinate dependent part. According to Ref.~\cite{Pol} such a term can produce coordinate-dependent translations. In other words it can curve the Dp-brane. We will assume that $A^D_i$ depends only on coordinates $x^i$ orthogonal to Dp-brane. So, in this paper we will work with a plane Dp-brane.

Let us now consider T-duality. In the present article we will work only with Abelian T-duality. When we use the formulation with field strength, according to Buscher's procedure \cite{B},  we will  gauge global shift symmetry. In the formulation with vector gauge fields we should apply the generalization of such procedure Ref.~\cite{DS2}.
Canonical momenta are T-dual  to the $\sigma$-derivative of the coordinates. After integration over $\sigma$ it turns to T-duality between momenta and winding numbers. On the other hand,  canonical  momenta are generators of the general coordinate transformations, while the $\sigma$-derivative of the coordinates are generators of the gauge symmetry \cite{EO,EO1,DS3}. It follows that general coordinate transformations are T-dual to the gauge symmetry, which is a fact  used in double field theories. In the open string case, after T-dualization additional vector fields with Neumann  boundary conditions turn into  vector fields with Dirichlet  boundary conditions, $A^N_a \to {}^\star A_D^b$, and vice versa, $A^D_i \to {}^\star A_N^j$.

We are going to carry through T-dualization in two ways: in terms of vector field and in terms of its field strength.  The first way is more challenging, because in that case the vector field is not constant and Buscher's procedure cannot be applied. The part with vector field which corresponds to the  Dirichlet  boundary conditions does not possess even global symmetry. So, we will use the
T-dualization procedure of Ref.~\cite{DS2}, which works in absence of global symmetry. We explicitly find  T-dual vector fields in the form
${}^\star {A}^a_D (V) = G_E^{-1 a b}  A_b^N (V) $ and $ {}^\star A^i_N (V) =  G^{-1 i j} {A}_j^D ( V) $. It shows that, as we expect, T-dualization changes boundary conditions and exchanges Neumann with Dirichlet
vector fields.  Additionally we prove that T-dual vector fields do not depend only on the dual coordinates $y_\mu$  but on $V^\mu$, which  besides $y_\mu$ depends  also on its double $\tilde y_\mu$.

The second way of T-dualization is simpler, because the field strength of the initial theory  is constant. The antisymmetric part ${\cal F}_{\mu \nu}^{(a)}$  can be considered an extension of the Kalb-Ramond field
while the symmetric part ${\cal F}_{\mu \nu}^{(s)}$  can be considered an extension of the metric tensor. So, it is easy to find complete T-dual background fields and T-dual field strength.

The   particular form of $V^\mu = - \kappa \, \theta^{\mu \nu} y_\nu + G^{-1 \mu \nu}_E \, {\tilde y}_\nu$  implies several features connected with non-geometric theories.  For example,
in Ref.~\cite{DNS2} it was shown that it produces non-associativity  of the coordinates, derived previously in the other way in Refs.~\cite{BHM,BMRS,DL,BDLPR,BP}.
In geometric theories the field strength for an Abelian vector field  is  simply $F_{\mu \nu}= \partial_\mu A_\nu - \partial_\nu A_\mu $. Because in non-geometric theories the vector field  depends on
$V^\mu$, we expect that T-dual  field strength will contain  derivatives  with respect to both variables $y_\mu$ and ${\tilde y}_\mu$.

The case with T-dual vector fields includes additional problems. The source of non-geometry is not only the argument $V^\mu$   of the vector background field but also the
T-dual $\sigma$-momentum  ${}^\star \gamma_{(0)}^\mu (y)$, which depends on both $\dot y_\mu$ and $y_\mu^\prime$. In that case we can analogously introduce
{\it T-dual effective vector fields} ${}^\star {\cal A}_{0}^\mu (V)$ and ${}^\star {\cal A}_{1}^\mu (V)$ in front of $\dot y_\mu$ and $y_\mu^\prime$  respectively  for both  Neumann and Dirichlet sectors.
T-duality allows us  to find  their dependence on the original vector fields $A_\mu$:
${}^\star {\cal A}_{\alpha}^\mu (A_\mu)\, ,\, (\alpha=0,1)$, which is equivalent to dependence on the original field strength ${}^\star {\cal A}_{\alpha}^\mu (F_{\mu \nu})\, ,\, (\alpha=0,1)$.

In the present article we will introduce the field strengths of non-geometric theories.
In geometric theories,   the term in the action with vector field (defined as integration over $\tau$) multiplied by $\dot x^\mu$   can be transformed to the term in  the action with corresponding field strength
(with integration over $d^2 \xi = d \tau d \sigma$). We can take this as a new definition for field strength. It agree with the standard one for geometric theories and provides us with new opportunities for
non-geometric theories.

We can generalize such an approach to the case of non-geometric theories. We will define  the effective T-dual field strength ${}^\star {\cal F}^{\mu \nu}$
as the term in action with integration over $d^2 \xi = d \tau d \sigma$
which is equivalent to the term with effective vector fields  ${}^\star {\cal A}_{0}^\mu (V)$ and ${}^\star {\cal A}_{1}^\mu (V)$
multiplied by  $\dot y_\mu$ and $y^\prime_\mu$ respectively. As well as in the initial theory,   besides the standard term antisymmetric in $\mu, \nu$ indices
${}^\star {\cal F}^{\mu \nu}_{(a)}= - {}^\star {\cal F}^{\nu \mu}_{(a)} $, it appears the new one is symmetric in $\mu, \nu$ indices
${}^\star {\cal F}^{\mu \nu}_{(s)}=  {}^\star {\cal F}^{\nu \mu}_{(s)} $.  The T-dual effective field strength depends on the initial one ${}^\star {\cal F}^{\mu \nu} (F_{\mu \nu})$. The expressions
${}^\star {\cal A}_{\alpha}^\mu (F_{\mu \nu})$ and ${}^\star {\cal F}^{\mu \nu} (F_{\mu \nu})$ allow us to eliminate the initial field strength  $F_{\mu \nu}$ and find expression
${}^\star {\cal F}^{\mu \nu}$ in terms of  ${}^\star {\cal A}_{\alpha}^\mu$.
In fact, first we can find  all antisymmetric and symmetric derivatives of ${}^\star {\cal A}_{\alpha}^\mu (F_{\mu \nu})$ with respect to both $y_\mu$ and ${\tilde y}_\mu$. Comparing these results with the known expression for
${}^\star {\cal F}^{\mu \nu} (F_{\mu \nu})$ we obtain the desired result.

Using the above results we will introduce genuinely non-geometric theories.
We will also discuss  local gauge symmetries of T-dual non-geometric theories as transformation of T-dual effective vector fields ${}^\star {\cal A}_{\alpha}^\mu (V)$ which does not change the T-dual
field strength ${}^\star {\cal F}^{\mu \nu}$. We will briefly discuss non-geometric matter fields.

%%%%%%%%%%%%%%%%%%%%%%%%%%%%%%%%%%%%%%%%%%%%%%%%%%%%%%%%%%%%%%%%%%%%%%%%%%%%%%%%%%%%%
\cleq

\section{T-duality of closed string}

In this section we introduce some known features of the bosonic closed string, which we are going to generalize to the case of the open string in the next sections. In particular, we consider  T-dual background fields and T-duality transformations in canonical form.

\subsection{Closed bosonic string}

Let us consider the closed bosonic string which propagates in D-dimensional space-time described by the action \cite{S}
\begin{equation}\label{eq:action0}
S[x] = \kappa \int_{\Sigma} d^2\xi\sqrt{-g}
\Big[\frac{1}{2}{g}^{\alpha\beta}G_{\mu\nu}[x]
+\frac{\epsilon^{\alpha\beta}}{\sqrt{-g}}B_{\mu\nu}[x]\Big]
\partial_{\alpha}x^{\mu}\partial_{\beta}x^{\nu},
\quad (\varepsilon^{01}=-1) \, .
\end{equation}

The string, with coordinates  $x^{\mu}(\xi),\ \mu=0,1,...,D-1$  is moving in a non-trivial background,
defined by the space metric $G_{\mu\nu}$ and the Kalb-Ramond field $B_{\mu\nu}$.
Here $g_{\alpha\beta}$ is the intrinsic world-sheet metric and
the integration goes over a two-dimensional world-sheet $\Sigma$
with coordinates $\xi^\alpha$ ($\xi^{0}=\tau,\ \xi^{1}=\sigma$).

Choosing the conformal gauge $g_{\alpha\beta}=e^{2F}\eta_{\alpha\beta}$, and introducing  light-cone coordinates
$\xi^{\pm}=\frac{1}{2}(\tau\pm\sigma)$, $\partial_{\pm}= \partial_{\tau}\pm\partial_{\sigma}$,
the action (\ref{eq:action0}) can be rewritten in the form
\begin{equation}\label{eq:action1}
S  = \kappa \int_{\Sigma} d^2\xi\
\partial_{+}x^{\mu}
\Pi_{+\mu\nu}
\partial_{-}x^{\nu},
\end{equation}
where we introduce a useful combination of background fields
\begin{eqnarray}
\Pi_{\pm\mu\nu} =
B_{\mu\nu} \pm\frac{1}{2}G_{\mu\nu}.
\end{eqnarray}

According to the action principle, variation  of the action (\ref{eq:action1})  with respect to $x^\mu$ produces an equation of motion
\begin{equation}\label{eq:eqmot}
\partial_{+}\partial_{-}x^\mu
+\Big{(}\Gamma^{\mu}_{\nu\rho}-B^{\mu}_{\ \nu\rho}\Big{)}
\partial_{+}x^\nu\partial_{-}x^\rho=0  \,  ,
\end{equation}
and boundary conditions
\begin{eqnarray}\label{eq:ibc}
\gamma^{(0)}_\mu (x) \delta x^\mu/_{\sigma=\pi} & - & \gamma^{(0)}_\mu  (x) \delta x^\mu/_{\sigma=0} =0 \,       ,
\end{eqnarray}
where $\Gamma^{\mu}_{\nu\rho}$ is the Christoffel symbol and   we introduce
\begin{equation}\label{eq:imp}
\gamma^{(0)}_\mu (x)  \equiv  \frac{\delta S}{\delta {x}^{\prime \mu}}=
\kappa\Big[2B_{\mu\nu} \dot{x}^\nu -G_{\mu\nu}{{x}}^{\prime \nu} \Big] =\kappa \Big( \Pi_{+ \mu \nu } \partial_- x^\nu + \Pi_{- \mu \nu } \partial_+ x^\nu   \Big)  \,  .
\end{equation}

The requirement of  world-sheet conformal invariance on the quantum level
leads to the space-time  equations of motion, which
at the lowest order in slope parameter $\alpha^\prime$,
for the constant dilaton field $\Phi=const$,  are
\begin{equation}\label{eq:ste}
R_{\mu \nu} - \frac{1}{4} B_{\mu \rho \sigma}
B_{\nu}^{\ \rho \sigma}=0\, ,  \qquad
D_\rho B^{\rho}_{\ \mu \nu} = 0   \, .
\end{equation}
Here
$B_{\mu\nu\rho}=\partial_\mu B_{\nu\rho}
+\partial_\nu B_{\rho\mu}+\partial_\rho B_{\mu\nu}$
is the field strength of the field $B_{\mu \nu}$, and
$R_{\mu \nu}$ and $D_\mu$ are the Ricci tensor and
covariant derivative with respect to the space-time metric.
We will consider the simplest  solutions of Eq.~(\ref{eq:ste}),
\begin{equation}
G_{\mu\nu}=const,
\quad
B_{\mu\nu}  = const  \,  ,
\end{equation}
which satisfy the space-time equations of motion.

\subsection{Sigma-model T-duality for closed string}

Applying the Buscher  T-dualization procedure along  all coordinates  \cite{B}, we obtain the T-dual action
\begin{equation}\label{eq:dualna}
{}^\star  S[y]=
\kappa
\int d^{2}\xi\
\partial_{+}y_\mu
\,^\star \Pi_{+}^{\mu\nu}  \,
\partial_{-}y_\nu
=\,
\frac{\kappa^{2}}{2}
\int d^{2}\xi\
\partial_{+}y_\mu
\theta_{-}^{\mu\nu}
\partial_{-}y_\nu,
\end{equation}
where
\begin{eqnarray}\label{eq:tpm}
{\theta}^{\mu\nu}_{\pm}&\equiv&
-\frac{2}{\kappa}
(G^{-1}_{E}\Pi_{\pm}G^{-1})^{\mu\nu}=
{\theta}^{\mu\nu}\mp \frac{1}{\kappa}(G_{E}^{-1})^{\mu\nu} \,  .
\end{eqnarray}
The symmetric and antisymmetric parts of ${\theta}^{\mu\nu}_{\pm}$  are the inverse of the effective metric $G^E_{\mu\nu}$ and the non-commutativity parameter $\theta^{\mu\nu}$
\begin{eqnarray}\label{eq:GEt}
G^E_{\mu\nu} & \equiv  G_{\mu\nu}-4(BG^{-1}B)_{\mu\nu}, \qquad
\theta^{\mu\nu} & \equiv  -\frac{2}{\kappa} (G^{-1}_{E}BG^{-1})^{\mu\nu}.
\end{eqnarray}

Consequently, the T-dual background fields  are
\begin{equation}\label{eq:tdbf}
^\star G^{\mu\nu} =
(G_{E}^{-1})^{\mu\nu}, \quad
^\star B^{\mu\nu} =
\frac{\kappa}{2}
{\theta}^{\mu\nu}  \, .
\end{equation}
Note that the dual effective metric is just the inverse of the initial metric
\begin{eqnarray}\label{eq:cdi1}
{}^\star G_E^{\mu\nu} \equiv {}^\star G^{\mu\nu}-4({}^\star B {}^\star G^{-1} {}^\star B)^{\mu\nu}= (G^{-1})^{\mu\nu}  \,  ,
\end{eqnarray}
and we will need the following relations
\begin{equation}\label{eq:cdi2}
({}^\star B {}^\star G^{-1})^\mu{}_\nu = - (G^{-1} B)^\mu{}_\nu  \,  , \qquad
({}^\star G^{-1} {}^\star B)_\mu{}^\nu = -(B G^{-1})_\mu{}^\nu  \,  .
\end{equation}

\subsection{T-duality transformations of closed string}

The T-duality transformation, connecting the variables $x^\mu$ of the initial closed string theory with  its corresponding T-dual ones $y_\mu$, takes the form \cite{DS1}
\begin{equation}\label{eq:xcong}
\partial_{\pm}x^\mu\cong
-\kappa\theta^{\mu\nu}_{\pm}
\partial_{\pm} y_\nu   \,  ,
\end{equation}
where the symbol $\cong$ denotes the T-duality relation.

From this equation we can find the T-dual transformation laws for $\dot{x}^\mu$ and $x^{\prime\mu}$,
\begin{eqnarray}\label{eq:dotx}
\dot{x}^\mu&\cong& -\kappa \theta^{\mu\nu} \dot y_\nu+(G_E^{-1})^{\mu\nu} y^\prime_\nu \, ,
\\\label{eq:xprime1}
x^{\prime\mu}&\cong& (G_E^{-1})^{\mu\nu} \dot y_\nu -\kappa\theta^{\mu\nu} y^\prime_\nu  \, .
\end{eqnarray}

It has been  shown in Ref. \cite{DS1} that the T-dual of the T-dual action is the original one.
The corresponding T-dual transformation  is the inverse of Eq.~(\ref{eq:xcong}),
\begin{equation}\label{eq:ct}
\partial_{\pm}y_\mu\cong
-2\Pi_{\mp\mu\nu}  \partial_{\pm}x^\nu     \, ,
\end{equation}
and consequently the transformation laws for $\dot{y}_\mu$ and $y^\prime_\mu$ are equal to
\begin{eqnarray}\label{eq:doty}
\dot{y}_\mu &\cong& -2B_{\mu\nu} \dot x^\nu +G_{\mu\nu}x^{\prime\nu}  \, ,
\\\label{eq:doty1}
y'_\mu &\cong&  G_{\mu\nu}\dot x^\nu-2B_{\mu\nu} x^{\prime\nu} \,  .
\end{eqnarray}

Using the expression for the canonical momentum of the original theory,
\begin{equation}\label{eq:mp}
\pi_\mu  \equiv \frac{\delta S}{\delta \dot{x}^\mu}=
\kappa\Big[G_{\mu\nu}{\dot{x}}^\nu-2B_{\mu\nu} x^{\prime\nu}\Big],
\end{equation}
and of the T-dual theory,
\begin{equation}\label{eq:dualmom}
^\star\pi^\mu
\equiv \frac{\delta\, ^\star S}{\delta \dot{y}_\mu}=
\kappa \Big[   (G^{-1}_{E})^{\mu\nu}
\dot{y}_\nu
-\kappa \theta^{\mu\nu}
y^\prime_\nu \Big]  \, ,
\end{equation}
we can rewrite the  transformations  (\ref{eq:xprime1})  and  (\ref{eq:doty1})   in the canonical form,
\begin{eqnarray}\label{eq:primex1}
\label{eq:xprime}
\kappa \, x^{\prime\mu} \cong
\,^\star\pi^\mu  \,  ,\qquad \pi_\mu \cong \kappa y^\prime_\mu   \, .
\end{eqnarray}
This relation connect momenta and winding numbers.

It was shown in Refs.~\cite{EO,EO1,DS3} that $\pi_\mu$ is the generator of general coordinate transformations while $x^{\prime \mu}$ is the generator of gauge symmetry. Then, Eq.~(\ref{eq:primex1}) shows that these symmetries are
T-dual to each other.

Since $\partial_\alpha x^\mu = \{ {\dot x}^\mu, x^{\prime \mu}  \} $ is a world-sheet vector, variation with respect to $\partial_\alpha x^\mu$,
\begin{eqnarray}\label{eq:wsv}
\pi^\alpha_\mu \equiv  \frac{\delta S}{\delta \partial_\alpha x^\mu }  = \{ \pi_\mu, \gamma^{(0)}_\mu (x)  \}           \, ,
\end{eqnarray}
is also a world-sheet vector. So,  the momentum $\pi_\mu$ and variable $\gamma^{(0)}_\mu (x)$, which will play  important roles in the analysis of boundary conditions, are components of the same world-sheet vector.
From now on we will call $\gamma^{(0)}_\mu (x)$ $\sigma$-momentum.

%%%%%%%%%%%%%%%%%%%%%%%%%%%%%%%%%%%%%%%%%%%%%%%%%%%%%%%%%%%%%%%%%%%%%%%%%%%%%%%%%%%%%
\cleq

\section{T-duality of open string}

In this section we will consider boundary conditions on the open string end-points  and adapt T-duality for such restrictions. Essentially,  all changes will happen on the string end points, although it is
useful to rewrite some expressions formally  as if they are  on the   world-sheet.

We will consider vector gauge field $A^N_a$ with Neumann boundary conditions,  which appears regularly in the literature. It is a $p+1$ dimensional vector on the Dp-brane.
It compensates the not-implemented gauge symmetry  of the Kalb-Ramond field  at the open string end-points.
In this article  we additionally introduce the $D-p-1$ dimensional vector field $A^D_i$  with Dirichlet boundary conditions, orthogonal to the Dp-brane, which with previous ones completes a $D$-dimensional vector.
It compensates the not-implemented general coordinate transformations at the open string end-points. We will show that field $A^D_i$ is T-dual to the $A^N_a$ one, as well as that the general coordinate transformations are T-dual to the
gauge symmetry of the Kalb-Ramond field.

\subsection{ T-duality between Dirichlet and Neumann    boundary conditions}

Unlike the closed string, the open string must satisfy boundary conditions at the string end-points.
For an initial    string they  take forms  (\ref{eq:ibc}) and  (\ref{eq:imp})  while for a T-dual string we have
\begin{eqnarray}\label{eq:tdbc}
^\star\gamma_{(0)}^\mu (y) \delta y_\mu/_{\sigma=\pi} &-& ^\star\gamma_{(0)}^\mu (y) \delta y_\mu/_{\sigma=0}   =0 \,       ,
\end{eqnarray}
where according to Eq.~(\ref{eq:dualna}) the T-dual $\sigma$-momentum is
\begin{equation}\label{eq:dualmom}
^\star\gamma_{(0)}^\mu (y)
\equiv \frac{\delta\, ^\star S}{\delta {y}_\mu^\prime}=
\kappa \Big[ \kappa \theta^{\mu\nu} \dot{y}_\nu  -  (G^{-1}_{E})^{\mu\nu} {y}_\nu^\prime  \Big] =
\kappa\Big[2 \,\,  {}^\star  B^{\mu\nu} \dot{y}_\nu - ^\star G^{\mu\nu} {{y}}^\prime_\nu \Big]
= \frac{\kappa^2}{2} \Big( \theta_-^{\mu \nu } \partial_- y_\nu + \theta_+ ^{\mu \nu } \partial_+ y_\nu   \Big)  \, .
\end{equation}
We can rewrite the T-dual transformations  (\ref{eq:dotx})  and (\ref{eq:doty})    in the form
\begin{eqnarray}\label{eq:dotx1}
\label{eq:xprime}
- \kappa \, \dot{x}^{\mu} \cong \,^\star\gamma^\mu_{(0)} (y) \,  ,\qquad    \gamma_\mu^{(0)} (x) \cong - \kappa \, \dot{y}_\mu   \, .
\end{eqnarray}

Note that we can put Eqs.~(\ref{eq:primex1}) and (\ref{eq:dotx1})  in compact, world-sheet covariant forms,
\begin{equation}\label{eq:wscf}
\kappa \partial_\alpha x^\mu \cong  \frac{\delta\, ^\star S}{\delta (\varepsilon^{\alpha \beta} \partial_\beta   y_\mu)}= - \varepsilon_{\alpha \beta} {}^\star \pi^{\beta \mu} \,  ,
\qquad
\kappa \partial_\alpha y_\mu \cong  \frac{\delta\, S}{\delta (\varepsilon^{\alpha \beta} \partial_\beta   x^\mu)}= - \varepsilon_{\alpha \beta} \pi^{\beta}_\mu   \,  .
\end{equation}

Let us show that the above equations connect the Dirichlet and Neumann boundary conditions.

1. If the end points (we will denote them with  $\partial \Sigma$ as a boundary of the world-sheet $\Sigma$) of the initial string satisfy  Neumann boundary conditions (which means that variation of some string end points
$\delta x^a/_{\partial \Sigma} \, $ with $a=0,1, \cdots , p$ is arbitrary) then $\gamma^{(0)}_a(x)/_{\partial \Sigma}=0 $.
Together with Eq.~(\ref{eq:dotx1}) it produces $\dot{y}_a /_{\partial \Sigma}=0$, which means that edges of the dual string are fixed. This is by definition the Dirichlet boundary conditions for a T-dual string.

2. Similarly, if the end points of an initial string satisfy  Dirichlet boundary conditions (it means that the edges of the string are fixed)  then $\dot{x}^i/_{\partial \Sigma}=0 \,$ where $i= p+1, \cdots , D-1$.
Together with  Eq.~(\ref{eq:dotx1}) it produces $ ^\star\gamma^i_{(0)}(y)/_{\partial \Sigma}=0$, which   according to Eq.~(\ref{eq:tdbc})   means that variations of the corresponding dual string end points  $\delta y_i/_{\partial \Sigma}$ are arbitrary. This is by definition the Neumann boundary conditions for a T-dual string.

\subsection{Neumann and Dirichlet vector background fields}

The action of closed string theory (\ref{eq:action0})  is invariant under local gauge transformations,
\begin{eqnarray}\label{eq:lgt}
\delta_\Lambda G_{\mu \nu} = 0 \, ,  \qquad  \delta_\Lambda B_{\mu \nu} = \partial_\mu \Lambda_\nu - \partial_\nu \Lambda_\mu \,  .
\end{eqnarray}
The open string theory is not invariant under these transformations. In Ref.~\cite{Zw} has been shown that for open strings,
\begin{eqnarray}\label{eq:dls}
\delta_\Lambda S[x] = 2\kappa \int d\tau (\Lambda_\mu {\dot x}^\mu/_{\sigma=\pi} -  \Lambda_\mu {\dot x}^\mu/_{\sigma=0}  ) \,  .
\end{eqnarray}

We already denoted the coordinates with  Neumann boundary conditions with $x^a$  and those with  Dirichlet boundary conditions with $x^i$. This means that $\delta x^a/_{\sigma=\pi}$ and $\delta x^a/_{\sigma=0}$
are  arbitrary, which produces $\gamma^{(0)}_a (x)/_{\sigma=\pi} = \gamma^{(0)}_a (x)/_{\sigma=0}=0$. On the other hand, $\delta x^i/_{\sigma=\pi}=0$ and $\delta x^i/_{\sigma=0} =0$, so that $\gamma^{(0)}_i (x)/_{\sigma=\pi}$  and
$\gamma^{(0)}_i (x)/_{\sigma=0}$  are arbitrary. So, because both string end points for $x^i$ coordinates satisfy Dirichlet boundary conditions, we have ${\dot x}^i/_{\sigma=0} = {\dot x}^i/_{\sigma=\pi}=0$ and consequently,
\begin{eqnarray}\label{eq:dlsN}
\delta_\Lambda S[x] = 2\kappa \int d\tau (\Lambda_a {\dot x}^a/_{\sigma=\pi} -  \Lambda_a {\dot x}^a/_{\sigma=0}  ) \,  .
\end{eqnarray}

To obtain gauge invariant action we should add the term
\begin{eqnarray}\label{eq:SA}
 S_{A_\Lambda} [x] = 2\kappa \int d\tau (A_a {\dot x}^a/_{\sigma=\pi} -  A_a {\dot x}^a/_{\sigma=0}  ) \,  ,
\end{eqnarray}
where the newly introduced vector field $A_a$  transforms with the same gauge parameter $\Lambda_a$,
\begin{eqnarray}\label{eq:dSA}
\delta_\Lambda   A_a  = - \Lambda_a      \,  .
\end{eqnarray}
Therefore, adding the term $S_{A_\Lambda} [x]$, we obtain the open string action invariant under local gauge transformations with parameter $\Lambda_a$.

It is natural to ask:

1. whether T-dual transformations of local gauge transformations exist (\ref{eq:lgt});

2. whether we can add some term $S_{A_\xi} [x]$ in order to obtain open string action invariant under such T-dual transformation;

3. whether the terms $S_{A_\Lambda} [x]$  and $S_{A_\xi} [x]$ are connected by T-duality  transformations as well as their origins.

In this article we will show that the answers to all these questions are affirmative. We can expect such a conclusion, because if T-duality is valid in the case of an open string then any step in the
original theory should have a partner in the T-dual version.

We expect that for every  characteristic of the initial theory we can find the corresponding one in T-dual theory.
In fact, we have the following table of related  terms in initial and   T-dual  theory:
\begin{eqnarray}
\begin{array}{ccccccc}
    G_{\mu\nu}           & B_{\mu\nu}                           &  - \kappa {\dot x}^a       &    \gamma^{(0)}_i (x) & LGT  &   A^N_a  (x)                &  A^D_i   (x)    \\
  {}^\star G^{\mu\nu}   & {}^\star B^{\mu\nu}     &   {}^\star \gamma_{(0)}^a (y) &   - \kappa {\dot y}_i   &  ? &   {}^\star  A_D^a (V)      &   {}^\star  A_N^i (V ) \\  \nonumber
\end{array}
\end{eqnarray}
where LGT is the abbreviation for  {\it local gauge transformations} and the question mark is for an unknown symmetry which we expect to be
``transformation T-dual to the local gauge transformation".
It will allow us to introduce the Dirichlet vector fields $A^D_i$  in analogy with the same procedure in which Neumann vector fields $A^N_a$ were introduced  in Ref.\cite{Zw}.
An interesting result has been obtained, that Dirichlet  and Neumann   vector fields, $A^D_i$  and $A^N_a$,  are  coupled to the T-dual expressions   $\gamma^{(0)}_i (x)$ and  $- \kappa {\dot x}^a$ respectively.
Later we will find that Dirichlet  and Neumann   vector fields are also T-dual to each other,
which   will complete the table above.

In Refs.~\cite{EO,EO1,DS3} it has been shown that  the generator of the local gauge transformations is $\sigma$-derivative of the coordinates $x^{\prime \mu }$.
Briefly, if the variation of the energy-momentum tensor $\delta \, T_\pm$ can be written as the Poisson bracket of some generator $\Gamma$ with energy-momentum tensor $T_\pm$, namely if the relation
\begin{eqnarray}\label{eq:emtPb}
\delta \, T_\pm = \{ \Gamma ,  T_\pm \}   \,  ,
\end{eqnarray}
is satisfied, then the corresponding transformation of background fields is the target-space  symmetry of the theory. For $\Gamma \to \Gamma_\Lambda = 2 \kappa \int d \sigma  \Lambda_\mu x^{\prime \mu }$, we just obtain  transformations (\ref{eq:lgt}).

According to Eq.~(\ref{eq:primex1}) the corresponding T-dual generator is $\Gamma_\xi = 2  \int d \sigma  \, \xi^\mu \pi_\mu$, with the following transformations of background fields:
\begin{eqnarray}\label{eq:gct}
\delta_\xi G_{\mu \nu} = -2 \, (D_\mu \xi_\nu + D_\nu \xi_\mu ) \, ,  \qquad
\delta_\xi B_{\mu \nu} =  -2 \, \xi^\rho B_{\rho \mu \nu} + 2 \partial_\mu (B_{\nu \rho} \xi^\rho) -  2 \partial_\nu (B_{\mu \rho} \xi^\rho)         \,  .
\end{eqnarray}
These transformations  have exactly the form of general coordinate transformations for background fields and they are  symmetry transformations of the space-time action.

Are they symmetries of the $\sigma$-model action?  The difference is transformation of the coordinates, which  does not appear explicitly  in target-space but  is present in $\sigma$-model action.
In fact, the energy-momentum tensor does not depend explicitly on the coordinates, and the above consideration does not give us information about  transformations of the coordinates.
In order to distinguish from standard general coordinate  transformations (which includes transformation of $x^\mu$), those  without transformation of $x^\mu$ we will call  transformations generated by $\pi_\mu$.

To better understand  what one  we have to choose as a T-dual to local gauge transformations in the $\sigma$-model action, it is useful to make transformations (\ref{eq:gct}) of the  background fields (metric tensor $G_{\mu \nu}$ and Kalb-Ramond field $B_{\mu \nu}$)    with parameter $\xi_\mu$ and the transformations of the string coordinates  $x^\mu$ with a different parameter   $\delta x^\mu = {\bar \xi}^\mu$. Then, using the equation of motion (\ref{eq:eqmot})   we obtain
\begin{eqnarray}\label{eq:otr}
\delta_\xi S[x] = - 2 \int d\tau \Big[ (\xi_\mu - {\bar \xi}_\mu)  \, G^{-1 \mu \nu} \gamma^{(0)}_\nu (x) /_{\sigma=\pi} -  (\xi_\mu - {\bar \xi}_\mu) \,  G^{-1 \mu \nu} \gamma^{(0)}_\nu (x) /_{\sigma=0}  \Big] \,  ,
\end{eqnarray}
where $\sigma$-momentum   $\gamma^{(0)}_\mu (x)$,  defined in Eq.~(\ref{eq:imp}),   is an expression which appears in the boundary conditions of the original theory.

First, we can conclude that at interior points of the string, on the equations of motions, the action is invariant even under separate transformations of background fields  and of the string coordinates.
For general coordinate transformations we have $\xi_\mu = {\bar \xi}_\mu$, and the whole action is invariant.
This is true even  without using equations of motion.
So, in the case of $\sigma$-model action for  an open string, we cannot accept reparametrization as a
T-dual of local gauge transformations. Such a choice does not allow us to add the corresponding vector fields, which should be T-dual to the fields $A_a$, introduced above.

Therefore, as a T-dual to local gauge transformations we will try to impose the  transformations (\ref{eq:gct}), the part of  general coordinate transformations,  which include  the transformations of background fields but do not include the transformations of the string coordinates  $x^\mu$.
Then we have ${\bar \xi}_\mu /_{\sigma=\pi} = {\bar \xi}_\mu /_{\sigma=0} =  0$ and
\begin{eqnarray}\label{eq:dlsgc}
\delta_\xi S[x] = - 2 \int d\tau \Big(\xi_\mu  \, G^{-1 \mu \nu} \gamma^{(0)}_\nu (x) /_{\sigma=\pi} -  \xi_\mu \,  G^{-1 \mu \nu} \gamma^{(0)}_\nu (x) /_{\sigma=0}  \Big) \,  .
\end{eqnarray}
Note that this  relation is a strong indication that  we are on the right track, because according to Eq.~(\ref{eq:dotx1})  $\dot x^\mu$ and $\gamma^{(0)}_\mu (x)$ are expressions T-dual to each other.
So, we obtained non-trivial transformations as we need and  the transformations (\ref{eq:lgt}) and (\ref{eq:gct}) are connected by T-duality.
From now on, for transformation (\ref{eq:gct})   we will use   expression: T-dual to local gauge transformations.

Let us for simplicity  assume that both metric tensor and Kalb-Ramond fields have a form
\begin{equation}
G_{\mu \nu} = \left (
\begin{array}{cc}
G_{a b}  &  0 \\
0  &   G_{i j}
\end{array}\right )\, ,  \qquad
B_{\mu \nu} = \left (
\begin{array}{cc}
B_{a b}  &  0 \\
0  &   B_{i j}
\end{array}\right ) \,  .
\end{equation}
Then for Neumann boundary conditions we have $\gamma^{(0)}_a (x)/_{\sigma=\pi} = \gamma^{(0)}_a (x)/_{\sigma=0}=0$, and thus  on
the equation of motion (\ref{eq:eqmot}) we are left  only with coordinates   that satisfy    Dirichlet boundary conditions,
\begin{eqnarray}\label{eq:dlsgc}
\delta_\xi S[x] = - 2 \int d\tau \Big(\xi_i  \, G^{-1 i j} \gamma^{(0)}_j (x) /_{\sigma=\pi} -  \xi_i \,  G^{-1 i j} \gamma^{(0)}_j (x) /_{\sigma=0}  \Big) \,  .
\end{eqnarray}

Consequently,  on the equations of motion the closed  string is invariant under transformations (\ref{eq:gct}),
while it is violated on the open string endpoints with Dirichlet boundary conditions. It remains to discuss the relation of transformations (\ref{eq:gct}) as a symmetry T-dual to the local gauge transformations, with
well known general coordinate transformations. The similarity is obvious, because the first transformations are  part of the second ones. There are two differences. One is a lack of transformation of the coordinates and the other is that transformations (\ref{eq:gct}) are symmetric only on the equations of motion. From the point of view of T-duality the second one is not a big surprise because, as it is well known, the equations of motion and Bianchi identity are
T-dual to each other.  So, local gauge transformations are symmetries  of the action without equations of motion, while its T-dual residual general coordinate transformations are  symmetries of the action on the equations of motion.

The next steps are similar to those in the case of local gauge transformations.
To obtain action invariant under residual general coordinate transformations  we should add the term
\begin{eqnarray}\label{eq:SAG}
 S_{A_\xi} [x] = - 2 \int d\tau \Big({A}_i G^{-1 i j}   \gamma^{(0)}_j (x) /_{\sigma=\pi} - {A}_i  G^{-1 i j} \gamma^{(0)}_j (x) /_{\sigma=0}  \Big) \,  ,
\end{eqnarray}
where the vector field ${A}_i $  transforms with the  gauge parameter of the residual general coordinate transformations $\xi_i$
\begin{eqnarray}\label{eq:dSAG}
\delta_\xi   {A}_i  = - \xi_i      \,  .
\end{eqnarray}Note that variation of $ S_{A_\xi} [x]$ does not include variation of metric $G^{-1  i j}$ and $\sigma$-momentum   $\gamma^{(0)}_j (x)$. In fact, ${A}_i$  is infinitesimal and variation of $G^{-1 i j}$ or $\gamma^{(0)}_j (x)$  will produce infinitesimals of the second order, which we will neglect.

So, the full  gauge invariant action for an open string is
\begin{eqnarray}\label{eq:giacom}
&S_{open} [x] =  S[x] + S_{A_\Lambda} [x] + S_{A_\xi} [x]     \\
&=  \kappa \int_{\Sigma} d^2\xi   \sqrt{-g}
\Big[\frac{1}{2}{g}^{\alpha\beta}G_{\mu\nu}[x]
+\frac{\epsilon^{\alpha\beta}}{\sqrt{-g}}B_{\mu\nu}[x]\Big]
\partial_{\alpha}x^{\mu}\partial_{\beta}x^{\nu}  \nonumber \\
&+  2\kappa \int d\tau \Big[ \Big( A_a^N [x] {\dot x}^a - \frac{1}{\kappa} {A}_i^D [x] G^{-1 i j}  \gamma^{(0)}_j (x) \Big)/_{\sigma=\pi} - \Big( A_a^N [x] {\dot x}^a - \frac{1}{\kappa} {A}_i^D [x] G^{-1 i j}
 \gamma^{(0)}_j (x) \Big)/_{\sigma=0} \Big] \,  . \nonumber
\end{eqnarray}
Consequently, the nontrivial background fields are $A_a \to A_a^N$ and ${A}_i \to A_i^D$, where we introduced the indices N and D for vector fields corresponding to Neumann and Dirichlet boundary conditions.

Note that  the  variables
\begin{eqnarray}\label{eq:phv}
{\cal B}_{a b} &= B_{a b} + \partial_a A_b^N - \partial_b A_a^N  \,  ,  \qquad     {\cal G}_{a b} = G_{a b}  \, ,  \nonumber\\
{\cal B}_{i j} &=   B_{i j} - 2 A^k_D  B_{k i j} + 2 \partial_i (B_{ j k} G^{-1 kq } A_q^D) -  2 \partial_j (B_{i k} G^{-1 kq } A_q^D)  \,  ,      \nonumber\\
{\cal G}_{i j} &= G_{i j} -2 ( \partial_i A_j^D + \partial_j A_i^D ) \,  ,
\end{eqnarray}
are gauge invariant under  Eq.~(\ref{eq:gct}), Eq.~(\ref{eq:lgt}) and transformations of the vector fields $\delta A_a^N = -  \Lambda_a$ and $\delta A_i^D = - \xi_i$,  and consequently they are  physical.
For further benefit let us introduce notations
\begin{eqnarray}\label{eq:FaFs}
 F^{(a)}_{a b}= \partial_a A_b^N - \partial_b A_a^N  \, ,     \qquad    F^{(s)}_{i j}= -2 ( \partial_i A_j^D + \partial_j A_i^D ) \, .
\end{eqnarray}

We are going to use the conformal gauge and the light-cone coordinates, so that the first term in $S_{open}$ obtains the form of the action (\ref{eq:action1}). For constant metric and Kalb-Ramond fields we have
\begin{eqnarray}\label{eq:Sopen}
& S_{open} [x] =  \kappa \int_{\Sigma} d^2\xi   \partial_{+}x^{\mu}      \Pi_{+ \mu\nu}  \partial_{-}x^{\nu}     \\
+&  2\kappa \int d\tau \Big[ \Big( A_a^N [x] {\dot x}^a - \frac{1}{\kappa} {A}_i^D [x] G^{-1 i j}  \gamma^{(0)}_j (x) \Big)/_{\sigma=\pi} - \Big( A_a^N [x] {\dot x}^a - \frac{1}{\kappa} {A}_i^D [x] G^{-1 i j}
 \gamma^{(0)}_j (x) \Big)/_{\sigma=0} \Big] \,  .  \nonumber
\end{eqnarray}

In the literature $A_a^N [x]$ is known as a massless vector field on the Dp-brane while ${A}_i^D [x]$ is known as massless scalar oscillations orthogonal to the Dp-brane. These are terms in relation to the Lorentz transformations that preserve the Dp-brane.

Note that inclusion of vector background fields changes the $\sigma$-momentum $\gamma^{(0)}_i$, defined in  Eq.~(\ref{eq:imp}). In fact, we will get an additional infinitesimal term  linear in the
vector background fields. It is multiplied by another infinitesimal, $A^D_i$, and consequently we will neglect it.

It is common to take both the  vector and the massless scalar fields to be constant, when the Buscher procedure can be applied. The constant massless scalar  field  performs uniform translation of the Dp-brane \cite{Pol}. We are going to use the generalized  procedure \cite{DS1,DS2}, so we are able to consider vector and massless scalar  fields linear in coordinates with infinitesimal coefficients. As explained in Ref.~\cite{Pol}, coordinate-dependent massless scalar   fields produce coordinate-dependent translations, which curve the Dp-brane. Consequently, our approach is able to describe an infinitesimally curved Dp-brane. We are not going to do this in the present
article, because for simplicity we will assume  later in Eq.~(\ref{eq:Alc})   that ${A}_i^D (x)$ depends only on $x^i$ coordinates and not on $x^a$.

\subsection{ T-dual background fields of open string}

Let us perform the T-dualization procedure on the theory described by the action (\ref{eq:Sopen}).  The first term contains constant background fields and so we can apply the standard Buscher's procedure of Section 2.2.
The remaining two terms are nontrivial because the background fields $A_a^N$ and ${A}_i^D$ are coordinate-dependent. To simplify the situation we will assume  that the vector fields are linear in coordinates,
\begin{eqnarray}\label{eq:Alc}
A_a^N (x)= A^0_a - \frac{1}{2} F_{a b}^{(a)} x^b \,  , \qquad  {A}_i^D (x) = A^0_i - \frac{1}{4} F_{i j}^{(s)} x^j \,   \,  ,
\end{eqnarray}
so that the corresponding field strengths   are constant. The coefficients  $F_{a b}^{(a)}$ and $F_{i j}^{(s)}$ are defined in Eq.~(\ref{eq:FaFs}).
The first coefficient is antisymmetric under $a,b$ indices while the second is symmetric under $i,j$ indices.

These forms of background fields satisfy the additional space-time equations of motion for open strings \cite{Le}. In our notation they take the form
\begin{eqnarray}\label{eq:osbf}
&  \beta_a = - \frac{1}{2} {\cal B}_a{}^b \partial_b \Phi + {\cal G}^{-1}_E {}^{b c}  \partial_c  {\cal B}_{b a} +
{\cal G}^{-1}_E {}^{b c} ( \frac{1}{2} {\cal B}_a{}^d B_{d b e} {\cal B}^e {}_c + K^\mu_{a c} B_{\mu \nu} \partial_b f^\nu )   \,  ,         \\
&  \beta_\mu =  \frac{1}{2} \partial_\mu \Phi   + {\cal G}^{-1}_E {}^{a b} ( \frac{1}{2} {\cal B}_b{}^c B_{\mu a c} -  K_{\mu a b})          \,  ,  \nonumber
\end{eqnarray}
where
\begin{eqnarray}\label{eq:dcbcg}
 {\cal B}_{a b} = B_{a b}  + F^{(a)}_{a b}  \,  ,   \qquad    {\cal G}^E_{a b} = G_{a b } - 4 {\cal B}_{a c}  G^{-1 c d}  {\cal B}_{d b} \,  ,
\end{eqnarray}
$B_{\mu \nu \rho}$  is the field strength of the Kalb-Ramond field $B_{\mu \nu}$ defined in  Section 2.1  and $K^\mu_{a b}$ is the extrinsic curvature. According to our assumptions $\Phi= const$ and ${\cal B}_{a b}= const$. So, $ {\cal G}^E_{a b}= const$ and $B_{a b c}=0$. Since we are working with a plane  Dp-brane
the extrinsic curvature is zero and both $\beta$-functions vanish.

Note that the part with Dirichlet vector field
\begin{eqnarray}\label{eq:Dvf}
& S_A^D [x] = - 2 \int d\tau \Big[ \Big(  {A}_i^D [x] G^{-1 i j}  \gamma^{(0)}_j (x) \Big)/_{\sigma=\pi} - \Big(  {A}_i^D [x] G^{-1 i j}
 \gamma^{(0)}_j (x) \Big)/_{\sigma=0} \Big]  \, \nonumber \\
& = 2 \kappa \int d\tau \Big[ \Big( 2  {\dot x}^i (B G^{-1})_i{}^j   {A}_j^D [x]  + x^{\prime i} {A}_i^D [x]  \Big)/_{\sigma=\pi} \nonumber  \\
& -  \Big( 2  {\dot x}^i (B G^{-1})_i{}^j   {A}_j^D [x]  + x^{\prime i} {A}_i^D [x] \Big)/_{\sigma=0} \Big] \,  ,
\end{eqnarray}
using the form of the vector field (\ref{eq:Alc}) after partial integration over $\tau$,  can be rewritten as
\begin{eqnarray}\label{eq:Dvfrw}
& S_A^D [x] =  2 \kappa \int d\tau \Big[ \Big( 2  {\dot x}^i  {A}_i^D [G^{-1} B x]  + x^{\prime i} {A}_i^D [x]  \Big)/_{\sigma=\pi} \nonumber  \\
& -  \Big( 2  {\dot x}^i  {A}_i^D [G^{-1} B x]  + x^{\prime i} {A}_i^D [ x] \Big)/_{\sigma=0} \Big] \,  .
\end{eqnarray}
So, we can conclude that following forms of the Dirichlet vector field are equivalent,
\begin{eqnarray}\label{eq:ADeq}
(B G^{-1})_i{}^j   {A}_j^D [x] \cong  {A}_i^D [G^{-1} B x]   \,  .
\end{eqnarray}

\subsubsection{ Auxiliary action}

Because parts with vector fields depend on the coordinate $x^\mu$ itself and not on  its derivatives with respect to $\tau$ and $\sigma$, it is not possible to apply the standard Buscher's procedure. So, we will need  generalized T-duality, developed in Ref.~\cite{DS1}.
Even more, the part with $A^D_i (x)$   does not have the global shift symmetry, because the expression $\gamma^{(0)}_i$ contains the part $G_{i j} \,  x^{\prime j}$  which is not the total derivative with respect to integration variable $\tau$.
So, we  should apply the T-dualization procedure of  Ref.~\cite{DS2}, which works in the absence of global symmetry.

Following  Ref.~\cite{DS2}, let us introduce the auxiliary action,
\begin{eqnarray}\label{eq:Soux}
& S_{aux} [v_\pm, y]  =   \kappa \int_{\Sigma} d^2\xi
\Big[ v_+^\mu  \Pi_{+\mu\nu} v_-^\nu    +\frac{1}{2} (v_+^\mu \partial_-y _\mu  - \partial_+y _\mu v_-^\mu )                   \Big]   \nonumber \\
&+  2\kappa \int d\tau \Big\{ \Big[ A_a^N (\Delta V) v_0^a - \frac{1}{\kappa} { A}_i^D (\Delta V) G^{-1 i j}   \gamma^{(0)}_j (V) \Big]/_{\sigma=\pi}  \nonumber \\
& - \Big[ A_a^N (\Delta V) v_0^a - \frac{1}{\kappa} { A}_i^D (\Delta V) G^{-1 i j}   \gamma^{(0)}_j (V) \Big]/_{\sigma=0} \Big\} \,  ,
\end{eqnarray}
where $\gamma^{(0)}_i (V) \equiv  \kappa (2 B_{ij} {\dot V}^j - G_{ij} {V^\prime}^j) =  \kappa (2 B_{ij} v_0^j - G_{ij} v_1^j)$ have been defined in accordance with Eq.~(\ref{eq:imp}).
It can be obtained from the  action (\ref{eq:Sopen}), by making  substitutions,
\begin{eqnarray}
\partial_{\pm} x^\mu\rightarrow v^{\mu}_{\pm} \,  , \quad   {\dot x}^\mu \rightarrow v^{\mu}_0   \,  ,     \quad    x^{\prime\mu} \rightarrow v^{\mu}_1     \,  ,    \quad             x^\mu\rightarrow\Delta V^\mu  \,  ,
\end{eqnarray}
and adding the Lagrange multiplier term with Lagrange multiplier  $y_\mu$. This action is constructed in the form of the gauge fixed action.
Here $v^{\mu}_{\pm}$ are some auxiliary fields, which take over the role of the gauge fields. Similarly to Refs.~\cite{DS1,DNS,DS2}, the argument of the background fields
is the line integral of the auxiliary fields
taken along a path $P$ (from $\xi_{0}$ to $\xi$),
\begin{equation}\label{eq:vdef}
\Delta V^\mu[v_{+},v_{-}]\equiv
\int_{P}d\xi^\alpha v^{\mu}_{\alpha}
=\int_{P}(d\xi^{+} v^{\mu}_{+}
+d\xi^{-} v^{\mu}_{-}).
\end{equation}

It is easy to show that the auxiliary action  $S_{aux} [x]$ (\ref{eq:Soux}) turns to the initial action $S_{open} [x]$  (\ref{eq:Sopen}).
Note that, also in Refs.~\cite{DS1,DS2}, the equation of motion with respect to $y_\mu$
forces the ``field strength" to vanish,
\begin{equation}\label{eq:emy}
\partial_{+}v^{\mu}_{-}
-\partial_{-}v^{\mu}_{+}=0 \, ,
\end{equation}
which is just the condition for the path independence of $\Delta V^\mu$. Using the solution of Eq.~(\ref{eq:emy}),
\begin{equation}\label{eq:vsol}
v^{\mu}_{\pm}=\partial_{\pm}x^\mu,
\end{equation}
one obtains $\Delta V^\mu(\xi)=x^\mu(\xi) - x^\mu(\xi_0)$, and  taking $x^\mu(\xi_{0})=0$
the auxiliary action reduces to the initial one (\ref{eq:Sopen}).

\subsubsection{ T-dual action}

The next step is to find the equations of motion with respect to  the auxiliary fields $v^\mu_{\pm}$. To prepare this, let us first rewrite the part of the action   (\ref{eq:Soux})
with integration over $d \tau$ to the integration over
$d^2 \xi = d \tau d \sigma$. We obtain
\begin{eqnarray}\label{eq:auxas}
&S_{aux} [v_\pm, y] =   \kappa \int_{\Sigma} d^2\xi
\Big[ v_+^\mu  \Pi_{+\mu\nu} v_-^\nu    +\frac{1}{2} (v_+^\mu \partial_-y _\mu  - \partial_+y _\mu v_-^\mu )                   \Big]   \\
+&  \kappa \int_{\Sigma} d^2\xi  \left\{ \Big[A_a^N (V) (v_+^a + v_-^a ) - 2 { A}_i^D (V)  G^{-1 i j} \Big( \Pi_{- j k} \, v_+^k + \Pi_{+ j k} \, v_-^k \Big) \Big] \Delta(\sigma) \right\} \,  , \nonumber
\end{eqnarray}
where $\Delta(\sigma) \equiv \delta(\sigma-\pi)- \delta(\sigma)$ and we used the relations
\begin{equation}\label{eq:vpm}
v^{\mu}_{\pm}=  v^{\mu}_0 \pm v^{\mu}_1  \,   ,
\end{equation}
and Eq.~(\ref{eq:imp}) for $\gamma^{(0)}_i (x)$.

Let us first calculate the variation over  the arguments  $V^\mu$ of the vector background fields. Using the form (\ref{eq:Alc})  of these fields and  zero order  equation of motion we can reexpress the term with vector fields
from Eq.~(\ref{eq:auxas}) in the form
\begin{equation}\label{eq:auxasv}
S_{A} [v_\pm, y] =
 \kappa \int_{\Sigma} d^2\xi  \left\{ \Big[A_a^N (V) (v_+^a + v_-^a ) + 2 { A}_i^D (G^{-1} \Pi_{+} V)  v_+^i + 2 { A}_i^D (G^{-1} \Pi_{-} V)  v_-^i   \Big] \Delta(\sigma) \right\} \,  .
\end{equation}
It helps us to find the variation with respect to arguments  $V^\mu$ of the  background fields,
\begin{eqnarray}\label{eq:deV}
& \delta_V S_{aux} [v_\pm, y]   \\
& =     \kappa \int_{\Sigma} d^2\xi  \left\{ (\delta v_+^a + \delta v_-^a ) A_a^N ( V) +
 2 \delta v_+^i  {A}_i^D ( G^{-1} \Pi_+  V)   + 2 \delta v_-^i {A}_i^D ( G^{-1} \Pi_- V)     \right\} \Delta(\sigma) \, . \nonumber
\end{eqnarray}
Now, the equations of motion after variation with respect to the auxiliary fields $v^\mu_{\mp}$ are:
\begin{equation}\label{eq:emv}
\Pi_{\mp\mu\nu} v_{\pm}^{\nu} +\frac{1}{2}\partial_{\pm}y_\mu  + \Big[ \mp 2 A_a^N ( V)   \mp 2 \Pi_{\mp i j}  G^{-1 j k} {A}_k^D ( V)    \mp  2 {A}_i^D ( G^{-1} \Pi_\mp V)  \Big] \Delta(\sigma)  = 0  \,  .
\end{equation}

Introducing new variables  ${\cal A}_{\pm \mu} (V) = \{{\cal A}_{\pm a} (V), {\cal A}_{\pm i} (V) \}$,
\begin{eqnarray}\label{eq:calA}
& {\cal A}_{\pm a} ( V) \equiv    A_{a}^N ( V)   \, , \nonumber \\
& {\cal A}_{\pm i} ( V) \equiv  \Pi_{\mp i j}  G^{-1 j k} {A}_k^D ( V)   + {A}_i^D ( G^{-1} \Pi_\mp V) \nonumber \\
&=  - \frac{1}{4} \left(B G^{-1} F^{(s)} + F^{(s)} G^{-1} B \mp  F^{(s)} \right)_{i j} V^j      \,  ,
\end{eqnarray}
or in components,
\begin{eqnarray}\label{eq:calAc}
& {\cal A}_{0 a} ( V) =   A_{a}^N ( V)   \, , \quad  {\cal A}_{1 a} ( V) = 0\,  , \\ \nonumber
& {\cal A}_{0 i} ( V) =  (B G^{-1})_i {}^j  A_j^D  ( V) +  A_i^D  (G^{-1} B V)                           = {\cal A}_{0 i}^{(0)}      - \frac{1}{4} (B G^{-1} F^{(s)} + F^{(s)} G^{-1} B)_{i j} V^j       \,  ,  \\ \nonumber
& {\cal A}_{1 i} ( V) = - A_i^{(0) D} +\frac{1}{4}  F_{i j}^{(s)} V^j  = - A_i^D  ( V)    \,  ,
\end{eqnarray}
we can rewrite the above equation as
\begin{equation}\label{eq:emvcA}
\Pi_{\mp\mu\nu}  v_{\pm}^{\nu} +\frac{1}{2}\partial_{\pm}y_\mu   \mp 2  {\cal A}_{\pm \mu}(V) \Delta(\sigma)  = 0  \,  .
\end{equation}

We introduced a pair of effective vector fields $ {\cal A}_{\alpha \mu} = \{  {\cal A}_{0 \mu} \, ,  {\cal A}_{1 \mu}  \}$ instead of the initial one $A_\mu = \{ A_{a}^N \, , A_{i}^D  \}$. So, we doubled the number of
vector fields, but there are two constraints on the effective vector fields,
\begin{eqnarray}\label{eq:cvf}
{\cal A}_{1 a} ( V)  =0 \, ,  \qquad     {\cal A}_{0 i} ( V) = - (B G^{-1})_i {}^j  {\cal A}_{1 j} ( V) - {\cal A}_{1 i} (G^{-1} B V)                 \,  .
\end{eqnarray}

The second constraint we can also rewrite in the forms
\begin{eqnarray}\label{eq:cvfr}
 (\Pi_+  G^{-1})_i {}^j   {\cal A}_{+ j} ( V) + {\cal A}_{+ i} (G^{-1} \Pi_+  V) =  (\Pi_-  G^{-1})_i {}^j   {\cal A}_{- j} ( V) + {\cal A}_{- i} (G^{-1} \Pi_-  V)    \,  .
\end{eqnarray}

Using Eq.~(\ref{eq:ADeq}), from now on  the $i$-components of Eqs.~(\ref{eq:calA}) and (\ref{eq:calAc})  we will  express  as
\begin{eqnarray}\label{eq:calAeq}
 {\cal A}_{\pm i} ( V) =  2 \Pi_{\mp i j}  G^{-1 j k} {A}_k^D ( V) \,  ,     \qquad  {\cal A}_{0 i} ( V) = 2 (B G^{-1})_i {}^j  A_j^D  ( V)     \,  .
\end{eqnarray}

Multiplying Eq.~(\ref{eq:emvcA}) from the left with $2 \kappa \theta_\pm$, we can solve it in terms of $v_\pm^\mu$,
\begin{equation}\label{eq:emvv}
 v_{\pm}^{\mu} = - \kappa \theta_\pm^{\mu \nu}  \partial_{\pm}y_\nu   \pm 4 \kappa \theta_\pm^{\mu \nu} {\cal A}_{\pm \nu} (V) \Delta(\sigma)
 =  - \kappa \theta_\pm^{\mu \nu} \Big(  \partial_{\pm}y_\nu   \mp 4  {\cal A}_{\pm \nu} (V) \Delta(\sigma) \Big) \,  ,
\end{equation}
or in components,
\begin{eqnarray}\label{eq:v0v1}
 v_{0}^{\mu} = - \kappa \theta^{\mu \nu} \Big[ {\dot y}_\nu -4 {\cal A}_{1 \nu} \Delta(\sigma)  \Big]   + G_E^{-1}{}^{\mu \nu}   \Big[  y^\prime_\nu -4 {\cal A}_{0 \nu} \Delta(\sigma)  \Big]  \,  , \\ \nonumber
 v_{1}^{\mu} = - \kappa \theta^{\mu \nu} \Big[  y^\prime_\nu -4 {\cal A}_{0 \nu} \Delta(\sigma)  \Big]   + G_E^{-1}{}^{\mu \nu}  \Big[ {\dot y}_\nu -4 {\cal A}_{1 \nu} \Delta(\sigma)  \Big]  \, .
\end{eqnarray}

Substituting  Eq.~(\ref{eq:emvv}) in Eq.~(\ref{eq:vdef})  we obtain
\begin{equation}\label{eq:Vsol1}
 V^{\mu} = - \kappa \theta^{\mu \nu} (y_\nu - 4 {\tilde {\cal A}}_{ \nu} )   + G^{-1}_E {}^{\mu \nu} ({\tilde y}_\nu  -4  {\cal A}_{ \nu} ) = V^{\mu} _0 + V^{\mu} _1   \,  ,
\end{equation}
where
\begin{eqnarray}\label{eq:AtA}
 \tilde {y}_{ \mu}   \equiv - \varepsilon_\alpha {}^\beta  \int d \xi^\alpha  \partial_\beta y_\mu
 = \int (d \tau y^\prime_{\mu}    + d \sigma  {\dot y}_{\mu}  )\, , \\ \nonumber
{\cal A}_{ \mu} \equiv \int d \xi^\alpha  {\cal A}_{\alpha \mu} \Delta(\sigma)   = \int (d \tau {\cal A}_{0 \mu}    + d \sigma  {\cal A}_{1 \mu}  ) \Delta(\sigma) \, , \\ \nonumber
 {\tilde {\cal A}}_{ \mu}   \equiv - \varepsilon_\alpha {}^\beta  \int d \xi^\alpha  {\cal A}_{\beta \mu} \Delta(\sigma)   = \int (d \tau {\cal A}_{1 \mu}    + d \sigma  {\cal A}_{0 \mu}  ) \Delta(\sigma)   \,  .
\end{eqnarray}

The  finite part $V^{\mu} _0$  and the infinitesimal one $V^{\mu} _1$ take a form
\begin{equation}\label{eq:V01}
 V^{\mu} _0  = - \kappa \theta^{\mu \nu} y_\nu    + G^{-1}_E {}^{\mu \nu} {\tilde y}_\nu  \, , \qquad
 V^{\mu} _1 =  4 \kappa \theta^{\mu \nu}  {\tilde {\cal A}}_{ \nu}   -4 G^{-1}_E {}^{\mu \nu}  {\cal A}_{ \nu}  \,  .
\end{equation}

We are going to substitute the solution   (\ref{eq:emvv})  back into the action (\ref{eq:auxas}). First we calculate
\begin{eqnarray}\label{eq:vpv}
  v_+^\mu  \Pi_{+\mu\nu} v_-^\nu & =& - \frac{\kappa}{2} \partial_{+}y_\mu \theta_{-}^{\mu\nu}  \partial_{-}y_\nu +2 \kappa \, {\cal A}_{+ \mu} (V)  \theta_{-}^{\mu\nu}  \partial_{-}y_\nu \Delta(\sigma)   \\ \nonumber
&- & 2 \kappa \partial_{+}y_\mu \theta_{-}^{\mu\nu}  {\cal A}_{- \nu} (V) \Delta(\sigma)
+ 8 \kappa  {\cal A}_{+ \mu} (V)  \theta_{-}^{\mu\nu}   {\cal A}_{- \nu} (V)  \Delta^2(\sigma) \,  ,
\end{eqnarray}
and
\begin{equation}\label{eq:gft}
\frac{1}{2} (v_+^\mu \partial_-y _\mu  - \partial_+y _\mu v_-^\mu )   = \kappa \partial_{+}y_\mu \theta_{-}^{\mu\nu}   \partial_{-}y_\nu -2 \kappa {\cal A}_{+ \mu} (V ) \theta_{-}^{\mu\nu} \partial_{-}y_\nu \Delta(\sigma)
 +2 \kappa \partial_{+}y_\mu \theta_{-}^{\mu\nu}  {\cal A}_{- \nu} (V)  \Delta(\sigma)  \,  .
\end{equation}
Consequently, the first part of the T-dual action is
\begin{equation}\label{eq:vpvgft}
  v_+^\mu  \Pi_{+\mu\nu} v_-^\nu + \frac{1}{2} (v_+^\mu \partial_-y _\mu  - \partial_+y _\mu v_-^\mu ) =  \frac{\kappa}{2} \partial_{+}y_\mu \theta_{-}^{\mu\nu}    \partial_{-}y_\nu
  + 8 \kappa  {\cal A}_{+ \mu} (V)  \theta_{-}^{\mu\nu}   {\cal A}_{- \nu} (V)  \Delta^2(\sigma)   \,  .
\end{equation}

Substituting the solution (\ref{eq:emvv})  into the part  of the action (\ref{eq:auxas}) with vector background fields, we have
\begin{eqnarray}\label{eq:avf}
& \left[ A_a^N (V) (v_+^a + v_-^a ) - 2 {A}_i^D (V) G^{-1 i  j}  B_{j k} (v_+^k + v_-^k ) + {A}_i^D (V) (v_+^i - v_-^i ) \right] \Delta(\sigma) \\ \nonumber
& = 2 \, \Big[ {A}_i^D (V) G^{-1 i  j}  {\dot y}_j - \frac{1}{\kappa} A_a^N (V) \,\,  {}^\star \gamma_{(0)}^a (y) \Big]  \Delta(\sigma) \\ \nonumber
& + 4 \kappa A^N_a (\theta_+^{a b} {\cal A}_{+ b} - \theta_-^{a b} {\cal A}_{- b}) \Delta^2 (\sigma)  -4 A^D_i G^{-1 i j} ({\cal A}_{+ j} - {\cal A}_{- j} ) \Delta^2 (\sigma) \,  ,
\end{eqnarray}
where ${}^\star \gamma_{(0)}^a (y)$ has been defined in Eq.~(\ref{eq:dualmom}). Similarly to the case of the initial theory, open string T-dual $\sigma$-momentum has an additional infinitesimal term proportional to the
T-dual vector fields. We will neglect it because it is multiplied by another infinitesimal $A_a^N (V) $.

Since the vector fields are infinitesimal, we can neglect   all  terms bilinear in the vector fields which helps us to avoid trouble with $\Delta^2(\sigma)$.  Consequently, the T-dual action takes the form
\begin{eqnarray}\label{eq:tdual}
& {}^\star S [y] =   \,
\frac{\kappa^{2}}{2}  \int d^{2}\xi\ \partial_{+}y_\mu \theta_{-}^{\mu\nu} \partial_{-}y_\nu  \nonumber \\
&+ 2 \kappa \int d \tau \Big[ \Big( {A}_i^D ( V) G^{-1 i  j}  {\dot y}_j - \frac{1}{\kappa} A_a^N (V) \, {}^\star \gamma_{(0)}^a (V)  \Big)/_{\sigma=\pi} \nonumber \\
&- \Big( {A}_i^D ( V) G^{-1 i  j}  {\dot y}_j - \frac{1}{\kappa} A_a^N (V) \, {}^\star \gamma_{(0)}^a   \Big)/_{\sigma=0}   (V)     \Big] \,  .
\end{eqnarray}

\subsubsection{T-dual background fields}

Because T-dual action should have the same form as the initial one (\ref{eq:Sopen}) but in terms of T-dual fields,
\begin{eqnarray}\label{eq:tdualf}
& {}^\star S [y] = \kappa \int d^{2}\xi  \partial_{+}y_\mu \,^\star \Pi_{+}^{\mu\nu}  \, \partial_{-}y_\nu  \nonumber \\
&+ 2 \kappa \int d \tau \Big[ \Big( {}^\star  A^i_N (V) {\dot y}_i - \frac{1}{\kappa} \, {}^\star {A}^a_D (V) \, {}^\star G^{-1}_{a b}  \, {}^\star \gamma_{(0)}^b (V)   \Big)/_{\sigma=\pi} \nonumber \\
&- \Big( {}^\star  A^i_N (V) {\dot y}_i - \frac{1}{\kappa} \, {}^\star {A}^a_D (V) \, {}^\star G^{-1}_{a b}  \, {}^\star \gamma_{(0)}^b (V)   \Big)/_{\sigma=0}  \Big] \,  ,
\end{eqnarray}
we can express  T-dual background fields in terms of the initial ones,
\begin{eqnarray}\label{eq:tdbfv}
^\star \Pi_{+}^{\mu\nu} = \frac{\kappa}{2} \theta_{-}^{\mu\nu}   \, , \quad   {}^\star {A}^a_D (V) = G_E^{-1 a b}  A_b^N (V)       \, , \quad  {}^\star A^i_N (V) =  G^{-1 i j} {A}_j^D ( V)     \, .
\end{eqnarray}
As one might expect, the T-dual metric and T-dual Kalb-Ramond fields remain the same as in the closed string case, Eq.~(\ref{eq:dualna}).

With the help of last two relations we can find effective T-dual vector fields in analogy with first relation in Eq.~(\ref{eq:calAeq}) and first relation in Eq.~(\ref{eq:calA}),
\begin{eqnarray}\label{eq:calAGd}
& {}^\star {\cal A}_{\pm }^a (V) = 2 \, {}^\star \Pi_{\mp}^{a b} \, {}^\star  G^{-1}_{b c} \, {}^\star {A}^c_D ( V)
= \kappa \, \theta_\pm^{a b} A_b^N  (V)   \, , \nonumber  \\
& {}^\star  {\cal A}_{\pm }^i ( V) =    {}^\star   A^i_N ( V)  =  G^{-1 i j} A_i^D  (V)   \,  .
\end{eqnarray}
In analogy with  second relation in Eq.~(\ref{eq:calAeq}), or from the previous relations, we have,
\begin{eqnarray}\label{eq:calAcd}
& {}^\star {\cal A}_{0 }^a ( V) = 2( {}^\star B \, {}^\star G^{-1})^a {}_b \,  {}^\star  A^b_D (V)  = \kappa \, \theta^{a b} A_b^N  (V)    \,  ,   \quad
 {}^\star {\cal A}_{1}^a ( V) =   - {}^\star A^a_D  ( V) = - G^{-1 a b}_E  A_b^N  (V)  \,  ,  \nonumber   \\
& {}^\star {\cal A}_{0}^i ( V) =  {}^\star A^i_N ( V)  = G^{-1 i j} A_j^D (V) \, ,  \qquad   {}^\star {\cal A}_{1}^i ( V) = 0   \,  .
\end{eqnarray}

We introduced two effective T-dual vector fields $ {}^\star {\cal A}_\alpha^\mu = \{ {}^\star  {\cal A}_0^\mu \, , {}^\star  {\cal A}_1^\mu  \}$ instead of the initial one
${}^\star A^\mu = \{{}^\star  A^{a}_D \, ,{}^\star  A^{i}_N  \}$, but  we have two constraints,
\begin{eqnarray}\label{eq:cvfd}
& {}^\star  {\cal A}_{0}^a ( V) =  - 2 ({}^\star B {}^\star G^{-1})^a{}_b  {}^\star {\cal A}_{1}^b ( V)
=  2 ( G^{-1} B)^a{}_b  {}^\star {\cal A}_{1}^b ( V)        \,  ,  \nonumber   \\
&  {}^\star {\cal A}_{1}^i ( V)  = 0   \, .
\end{eqnarray}
The first relation we can rewrite in the forms
\begin{eqnarray}\label{eq:cvfrr}
 {}^\star \Pi_{+}^{a b}  {}^\star G^{-1}_{b c}    {}^\star  {\cal A}_{+}^c ( V)
 =  {}^\star \Pi_{-}^{a b}  {}^\star G^{-1}_{b c}    {}^\star  {\cal A}_{-}^c ( V)     \, , \qquad
  \Pi_{- a b}  {}^\star  {\cal A}_{+}^b ( V)
=   \Pi_{+ a b}  {}^\star  {\cal A}_{-}^b ( V)    \,  .
\end{eqnarray}

Let us make two observations. First, vector fields corresponding to Neumann (Dirichlet)  boundary conditions of the initial theory $A_a^N (x)$ in front of ${\dot x}^a$ in  Eq.~(\ref{eq:Sopen})
(${A}_i^D (x)$ in front of $\gamma^{(0)}_i$ in  Eq.~(\ref{eq:Sopen})) after T-dualization  turn to the fields corresponding to
Dirichlet  (Neumann) boundary conditions of the T-dual theory ${}^\star {A}^a_D (V)$ in front of ${}^\star \gamma_{(0)}^a$ in  Eq.~(\ref{eq:tdualf})
($ {}^\star  A^i_N (V)$ in front of ${\dot y}_i$ in  Eq.~(\ref{eq:tdualf})).  Therefore, T-duality interchanges Neumann with Dirichlet gauge fields.
Second, the T-dual vector background fields depend not on $y_\mu$ but on the finite part of  Eq.~(\ref{eq:V01}),
\begin{eqnarray}\label{eq:solV}
V^\mu \to  V_0^\mu= - \kappa \, \theta^{\mu \nu} y_\nu + G^{-1 \mu \nu}_E \, {\tilde y}_\nu   \, .
\end{eqnarray}
We can neglect the infinitesimal part $V_1^\mu$  because it always appears in the argument of the vector background fields, with an infinitesimal coefficient. So it will
produce the square of the vector fields, which we will neglect.

The variable ${\tilde y_\mu}$ naturally appears in Buscher's approach, as a part of variable $V^\mu$   when we perform T-dualization
along coordinates on which background fields depend. Then we must introduce gauge invariant coordinates which are line integrals of the covariant
derivatives. The corresponding argument of T-dual background fields  $V^\mu$  is a solution of T-dual transformation  laws and depends not only on $y_\mu$, but is a linear combination of both $y_\mu$ and ${\tilde y_\mu}$.
Therefore, the variable $V^\mu (y_\mu , {\tilde y})$, and not  variable $y_\mu$, is T-dual to $x^\mu$.

The variable ${\tilde y_\mu}$ is defined in terms of $y_\mu$, see Eq.~(3.45), as a line integral of $\sigma$ and
$\tau$ derivatives of  $y_\mu$. It produces non-locality of the arguments of background fields, which in our formulation is the source of non-geometry.
On the finite part of the equation of motion (a case that  always happens)  it does not depend on the integration path.
Because it always  appears as a part of variable $V^\mu$,  we can take  for it  the same boundary conditions as for variable $y_\mu$. Then the variable $V^\mu$ has definite boundary conditions.

The variable ${\tilde y_\mu}$, as a part of  $V^\mu$, is significant because it distinguishes non-geometric from  geometric theories.
In the literature, these kinds of theories are recognized as  theories with R-flux. Some authors  refer to them as exotic configurations. I expect that just background field dependence on $V^\mu (y_\mu , {\tilde y}_\mu)$ is the source of these exotic non-geometric behavior. In fact, as was shown in Ref.~[5], the presence of ${\tilde y_\mu}$ produces non-commutativity of the closed string variables and non-associativity.

Later, in Sections 5 and 6, we will see that ${\tilde y_\mu}$ has a central role in the definition of field strength for non-geometric theories, and is a basic variable for truly non-geometric theories.

Note that the T-dual of the T-dual produces the initial background fields. For example,
\begin{eqnarray}\label{eq:tdtd}
^\star {}^\star  A_a^N (x)  =  {}^\star G^{-1}_{a b}  {}^\star {A}^b_D (V) = G^E_{a b}  G_E^{-1 b c}   A_c^N (x) = A_a^N (x)     \, , \\ \nonumber
^\star {}^\star  A_i^D (x)  =  {}^\star G^{-1}_{E i j}  {}^\star {A}^j_N (V) = G_{i j}  G^{-1 j k}   A_k^D (x) = A_i^D (x)     \, .
\end{eqnarray}

From Eqs.~(\ref{eq:vsol})  and  (\ref{eq:emvv})  we can find the T-dual transformation law,
\begin{equation}\label{eq:tdtr}
 \partial_{\pm} x^{\mu}  \cong   - \kappa \theta_\pm^{\mu \nu}  \partial_{\pm}y_\nu   \pm 4 \kappa \theta_\pm^{\mu \nu} {\cal A}_{\pm \nu} (V) \Delta(\sigma) \,  ,
\end{equation}
while  its inverse is
\begin{equation}\label{eq:itdtr}
\partial_{\pm}y_\mu   \cong  -2 \Pi_{\mp \mu \nu} \partial_{\pm} x^{\nu}    \pm  4   {\cal A}_{\pm \mu} (x)  \Delta(\sigma) \,  .
\end{equation}
In fact the last transformation can be obtained after T-dualization of the T-dual action (\ref{eq:tdual}).
Both transformations differ from the closed string ones by the infinitesimal term which contains vector background fields ${\cal A}_{\pm \mu}$.

\subsection{Relation with standard approach}

There are significant differences between present and standard T-duality transformations of the open string. First, we are working with constant field strength (gauge field linear in
compactified coordinates) while in the  standard approach the field strength is zero (gauge field is independent of compactified coordinates). Second, and most important, in the present article both Neumann and Dirichlet gauge fields are introduced through the boundary coupling in the action: the Neumann one through  coupling  with ${\dot x}^a$ and the  Dirichlet one through  coupling  with $\gamma^{(0)}_i$. As a difference of the standard approach they are treated in the same way.
The Lagrangian treatment of Dirichlet gauge fields through the term
$A^D_i G^{-1 i j} \gamma_i^{(0)}$  has not previously been presented  in the literature.  We will see that
the problem with the standard approach is that it misses such a  Dirichlet part in the action. Third, which is a more technical difference, in the standard approach T-duality has been performed along one direction while in our approach it is performed along an arbitrary set of directions.

Let us start with the choice of auxiliary action. For discussion of this subsection it is useful to introduce T-dual coordinates $y_a$ through the part of the action
\begin{eqnarray}\label{eq:caa}
\Delta S_{aux}  =   \frac{\kappa}{2}  \int_{\Sigma} d^2\xi \,  y _a F^a_{+ -}   =   \frac{\kappa}{2}  \int_{\Sigma} d^2\xi \,  y _a
(\partial_+ v_-^a - \partial_- v_+^a )       \,  ,
\end{eqnarray}
which can be reexpressed as
\begin{eqnarray}\label{eq:caar}
\Delta S_{aux} =  \frac{\kappa}{2}  \int_{\Sigma} d^2\xi (v_+^a \partial_- y _a  - \partial_+ y _a v_-^a ) +  \kappa \int_{\partial \Sigma} d \tau  \,  y _a \,  v_0^a           \,  .
\end{eqnarray}
The last term changes boundary part  (\ref{eq:auxasv}) in such a way that
\begin{eqnarray}\label{eq:Anpy}
A^N_a \to A^N_a + \frac{1}{2} \, y_a  \,  .
\end{eqnarray}
Note that according to the relation $\delta A^N_a = - \Lambda_a$, this additional term is just a gauge transformation with $\Lambda_a = - \frac{1}{2}y_a $. So, up to  gauge transformation this choice of auxiliary action
is equivalent to the previous one  (\ref{eq:Soux}).

Let us for the sake of discussion preserve this boundary  term (not gauge it away).
To understand the relationship between ours and previous  approaches \cite{DO,ABB}, let us reduce our case to the standard one. We will suppose that the field strengths for both Neumann and Dirichlet gauge fields are zero,
which means that these fields are constant.  Finally, because this is the first time Lagrangian treatment of Dirichlet gauge fields has been performed, we should  put the Dirichlet field of the initial action equal to zero, $A^D_i =  0$.

The T-dual action must  have the same form as the initial one but in terms of T-dual fields.
According to Eq.~(\ref{eq:tdbfv}) and taking into account the new term  (\ref{eq:Anpy}),  we obtain an expression for T-dual vector fields,
\begin{eqnarray}\label{eq:tdbfv1}
{}^\star {A}^a_D  = G_E^{-1 a b} \left( A_b^N  + \frac{1}{2} y_a \right)      \, , \quad  {}^\star A^i_N  =  G^{-1 i j} {A}_j^D     \, .
\end{eqnarray}
Because the standard approach  started with $A^N_a=const$ and $A^D_i =  0$, it produces
\begin{eqnarray}\label{eq:tdbfv2}
{}^\star {A}^a_D  = G_E^{-1 a b}  ( A_b^N + \frac{1}{2} y_a)         \, , \quad  {}^\star A^i_N  =  0     \, .
\end{eqnarray}
Let us stress that in the standard approach,  T-dual action does not contain  either Neumann  or Dirichlet  vector fields. Generally, a term with T-dual Neumann fields ${}^\star A^i_N $  could be recognized as a coefficient in  front of ${\dot y}_i$, but according to the last relation it is zero. This is a consequence of the fact that
one did not know how to include initial Dirichlet fields and  had to  put it to zero. It follows that the T-dual Neumann field is also zero.

In the standard approach one cannot recognize the T-dual Dirichlet vector field
${}^\star {A}^a_D$,  which according to the present paper should be in front of $^\star\gamma_{(0)}^\mu (y)$. So, for consistency of the standard approach, one should require that it vanishes, ${}^\star {A}^a_D=0$.
According to  Eq.~(\ref{eq:tdbfv2}),  this is in fact the Dirichlet boundary condition  of the standard approach, $y_a = - 2 A^N_a$.

Consequently, the problem of the standard approach,  which has been solved in the present article,  is ignorance in introducing the Dirichlet background field in both initial and T-dual Lagrangians.
It means that the consistency of the standard approach ${}^\star {A}^a_D=0$ produces $y_a = - 2 A^N_a$.   This has the interpretation  of Dirichlet boundary conditions for  T-dual theory. This is an external condition which  has not been obtained  from the Lagrangian. If we   perform T-duality along one direction (let us say $a=1$), than we obtain the well known result of the standard approach,
$y = - 2 A^N$ .

%%%%%%%%%%%%%%%%%%%%%%%%%%%%%%%%%%%%%%%%%%%%%%%%%%%%%%%%%%%%%%%%%%%%%%%%%%%%%%%%%%%%%
\cleq

\section{ T-duality in terms of field strengths}

In the previous section  we investigated the T-duality of the vector fields. In the initial (geometric) theory we  considered gauge fields linear in coordinates $x^\mu$. We obtained that the gauge fields of the T-dual (non-geometric)
theory are linear in the new variable  $V^\mu$, which is a function of the T-dual coordinate $y_\mu$ and its double ${\tilde y}_\mu$.

Generally, it is not clear how to define the field strength for non-geometric theories. So we will go a roundabout way.
It is known that in geometric theories,  if both ends of the open string are attached to the same Dp-brane,  the term in the action which contains the vector background field with  integration over $\tau$
can be transformed to the term in  the action which contains  corresponding field strength with integration over $d^2 \xi = d \tau d \sigma$. We are going to generalize such a relation to non-geometric theories.

\subsection{Field strengths of initial theory}

After some direct calculation for Neumann vector fields we obtain
\begin{eqnarray}\label{eq:ttots0}
&S_A^N [x] =  2\kappa \int d\tau \Big[ \Big( A_a^N [x] {\dot x}^a  \Big)/_{\sigma=\pi} -
\Big( A_a^N [x] {\dot x}^a  \Big)/_{\sigma=0} \Big] \,  , \nonumber\\
& =  -2 \kappa \int d^2  \xi  {\dot x}^a {\cal F}_{a b} \, x^{\prime b}  =  \kappa \int d^2  \xi \partial_+  x^a {\cal F}_{a b} \, \partial_- x^b  \,  ,
\end{eqnarray}
where only the antisymmetric part contributes,
\begin{equation}\label{eq:FcF0}
{\cal  F}_{a b}  =  {\cal  F}_{a b}^{(a)} = F^{(a)}_{a b}  = \partial_a A_b^N (x) - \partial_b A_a^N (x)    =  \partial_a {\cal A}_{0 b} (x) - \partial_b {\cal A}_{0 a} (x)
  \,  ,
\qquad   {\cal  F}_{a b}^{(s)}  = 0  \, .
\end{equation}

We can trivially reexpress Eq.~(\ref{eq:ttots0})    in a  form where the effective background vector field  ${\cal A}_{0 a}$ is multiplied with ${\dot x}^a$,
\begin{eqnarray}\label{eq:cAdx}
S_A^N [x] =  2\kappa \int d\tau \Big( {\cal A}_{0 a} [x] {\dot x}^a/_{\sigma=\pi}  - {\cal A}_{0 a} [x] {\dot x}^a/_{\sigma=0}     \Big) \, .
\end{eqnarray}

We introduce the names {\it effective background vector field} ${\cal A}_{ a}$  and corresponding {\it effective field strength} ${\cal  F}_{a b} $ for variables obtained in this way. In this simplest case
we have a standard picture: one  effective vector field ${\cal A}_{0 a} = A_a^N$ and corresponding  antisymmetric effective  field strength ${\cal  F}_{a b}  = {\cal  F}_{a b}^{(a)} $.  In the next cases the situation will be more complicated.

Instead of Eq.~(\ref{eq:FcF0}) we can also accept the relation  (\ref{eq:ttots0})  as definitions of the field strength for geometric theories. Let us
extend this definition to  non-standard  theories.
Unlike Neumann vector fields, which are coupled with $\dot x^a$, Dirichlet vector fields are coupled with $\sigma$-momentum  $\gamma^{(0)}_j (x)$, which will produce additional problems.
Using  the finite part of the equation of motion  $ {\ddot{x}}^i = x''^i$,  we have
\begin{eqnarray}\label{eq:ttots}
&S_A^D [x] =  2\kappa \int d\tau \Big[ \Big( - \frac{1}{\kappa} {A}_i^D [x]  G^{-1 i j}  \gamma^{(0)}_j (x) \Big)/_{\sigma=\pi} -
\Big( - \frac{1}{\kappa} {A}_i^D [x]  G^{-1 i j}  \gamma^{(0)}_j (x) \Big)/_{\sigma=0} \Big] \,  , \nonumber\\
& =  \kappa \int d^2  \xi  \, \partial_+  x^i \,  {\cal F}_{i j} \, \partial_- x^j     \,  .
\end{eqnarray}
Now, both symmetric and antisymmetric parts contribute
\begin{equation}\label{eq:Fij}
{\cal F}_{i j} = {\cal F}_{i j}^{(a)} + \frac{1}{2} {\cal F}_{i j}^{(s)} \, ,
\end{equation}
where
\begin{eqnarray}\label{eq:FcF}
& {\cal F}_{i j}^{(a)} = \Big[ \partial_i \Big(2 B_{j k} G^{-1 k q} {A}^D_q \Big) - \partial_j \Big(2 B_{i k} G^{-1 k q} {A}^D_q \Big)   \Big]
=      \frac{1}{2}  \Big( B_{i k} G^{-1 k q } F^{(s)}_{q j} + F^{(s)}_{i k} G^{-1 k q}B_{q j} \Big)  \nonumber\\
& =  F_{i j}^{(a)}   = \partial_i {\cal A}_{0 j} (x) - \partial_j {\cal A}_{0 i} (x) \,   \,     ,
\end{eqnarray}
and
\begin{eqnarray}\label{eq:FcFs}
{\cal F}_{i j}^{(s)} = -2 (\partial_i A_j^D + \partial_j A_i^D ) =  {F}_{i j}^{(s)}  = 2 \Big(  \partial_i {\cal A}_{1 j} (x) + \partial_j {\cal A}_{1 i} (x) \Big) \,   .
\end{eqnarray}

For the Dirichlet sector, an analogy with the standard approach  does not exist. In that case both components of the effective background vector field,   ${\cal A}_{0 i}$ and ${\cal A}_{1 i}$, as well as both ${\dot x}^i$ and ${x}^{\prime i}$, contribute. So, we  can reexpress  Eq.~(\ref{eq:ttots}) as
\begin{eqnarray}\label{eq:cAdxd}
S_A^D [x] &=2\kappa \int d\tau \Big[ \Big( {\cal A}_{0 i} [x] {\dot x}^i - {\cal A}_{1 i} [x] {x}^{\prime i} \Big)/_{\sigma=\pi} - \Big(  {\cal A}_{0 i} [x] {\dot x}^i - {\cal A}_{1 i} [x] {x}^{\prime i} \Big)/_{\sigma=0}  \Big] \, , \nonumber\\
&  =   2\kappa  \eta^{\alpha \beta} \int d\tau  \Big(    {\cal A}_{\alpha i} [x] \partial_\beta  {x}^i/_{\sigma=\pi} - {\cal A}_{\alpha i} [x] \partial_\beta  {x}^i/_{\sigma=0} \Big)  \nonumber\\
& =   2\kappa   \int d\tau  \Big(    {\cal A}_{\alpha i} [x]   {\dot x}^{\alpha i}/_{\sigma=\pi} - {\cal A}_{\alpha i} [x]  {\dot x}^{\alpha i}/_{\sigma=0} \Big)  \,  ,
\end{eqnarray}
where ${\cal A}_{0 i} [x]$ has been defined in  Eq.~(\ref{eq:calAeq}),  ${\cal A}_{1 i} [x]$ in Eq.~(\ref{eq:calAc}), and we introduced the notation
$ {\dot x}^{\alpha i}= \{{\dot x}^i , -{ x}^{\prime i} \}= \eta^{\alpha \beta} \partial_\beta x^i$.

Note that,  although we work with an initial  theory,   this action does depend on ${\cal A}_{1 i} [x]$ and the vector field couples not only with $\dot x^i$ but also with $x^{\prime i}$. This is consequence of the fact that the original vector field ${A}_i^D (x)$ is   not multiplied by ${\dot x}^i$ but by $\sigma$-momentum   $G^{-1 i j}  \gamma_j^{(0)} (x)$.

\subsection{Field strengths of T-dual theory}

The case with T-dual theory is  more complicated because the vector fields depend on $V^\mu$, which is a function of two variables,
$y_\mu$ and ${\tilde y}_\mu$.

\subsubsection{The case of Dirichlet vector fields}

For Dirichlet vector fields,  with the help of the finite part of the equation of motion  $ {\ddot{y}}_a = y''_a$,    we find
\begin{eqnarray}\label{eq:dttots}
& ^\star S_A^D [y] =  2 \kappa \int d \tau \Big[ \Big(  - \frac{1}{\kappa}  {}^\star {A}^a_D (V) \, {}^\star G^{-1}_{a b}  \, {}^\star \gamma_{(0)}^b (y)  \Big)/_{\sigma=\pi} -
\Big( - \frac{1}{\kappa} {}^\star {A}^a_D (V) \, {}^\star G^{-1}_{a b}  \, {}^\star \gamma_{(0)}^b  (y) \Big)/_{\sigma=0}  \Big]   \,  , \nonumber\\
& =  \kappa \int d^2  \xi  \partial_+  y_a \, {}^\star {\cal F}^{a b} \, \partial_- y_b  \,  .
\end{eqnarray}
Here we have
\begin{equation}\label{eq:Fab}
{}^\star {\cal F}^{a b} = {}^\star {\cal F}^{a b}_{(a)} + \frac{1}{2} \, {}^\star {\cal F}^{a b}_{(s)} \, ,
\end{equation}
where the antisymmetric part,
\begin{eqnarray}\label{eq:dFcFD}
^\star {\cal F}^{a b}_{(a)} =  \kappa \left( {}^\star G_E \, {}^\star \theta \, {}^\star  F \, {}^\star B - \, {}^\star B \, {}^\star F^T \, {}^\star \theta \, {}^\star  G_E   \right)^{a b} -
\frac{1}{2} ( {}^\star F^a {}_c \, {}^\star G^{c b} - {}^\star G^{a c} \, {}^\star F^T {}_c {}^b )   \nonumber\\
= 2 \, {}^\star B^{a c} ( {}^\star F^T  {}^\star G^{-1} - {}^\star G^{-1}  {}^\star F)_{c d} {}^\star B^{d b}  -
\frac{1}{2} ( {}^\star F^a {}_c \, {}^\star G^{c b} - {}^\star G^{a c} \, {}^\star F^T {}_c {}^b )       \, ,
\end{eqnarray}
and symmetric part,
\begin{eqnarray}\label{eq:dFcFDs}
^\star {\cal F}^{a b}_{(s)} =  -4  \left(  {}^\star  F \, {}^\star B  +  {}^\star B \, {}^\star  G^{-1}  \, {}^\star F \, {}^\star  G   \right)^{a b}     \, ,
\end{eqnarray}
are expressed in terms of coefficient ${}^\star F^a {}_b$, defined with the relation
\begin{eqnarray}\label{eq:sAD}
^\star A^a_D (V) = {}^\star A^a_0  - \frac{1}{2} \,  {}^\star F^a {}_b V^b    \, .
\end{eqnarray}

Taking into account that with the help of Eq.~(\ref{eq:tdbfv}),  from  the first relation  in Eq.~(\ref{eq:Alc}) and  Eq.~(\ref{eq:sAD})  we have ${}^\star F^a {}_b = G^{-1 a c}_E F_{c b}$, and  it follows that:
\begin{eqnarray}\label{eq:dFcFDi}
^\star {\cal F}^{a b}_{(a)} =   - \kappa^2 \theta^{a c}  {F}_{c d}^{(a)}  \, \theta^{d b} - G_E^{-1 ac}  {F}_{c d}^{(a)}  \, G_E^{-1 d b}
= - \frac{\kappa^2}{2} \Big[ \theta_+^{a c}  { F}_{c d}^{(a)}  \theta_+^{d b}  + \theta_-^{a c}  {F}_{c d}^{(a)}  \theta_-^{d b}  \Big]       \, ,
\end{eqnarray}
and
\begin{eqnarray}\label{eq:dFcFDis}
^\star {\cal F}^{a b}_{(s)} =   - 2 \kappa \Big[  G_E^{-1 a c}  {F}_{c d}^{(a)}  \, \theta^{d b} +  \theta^{a c}   {F}_{c d}^{(a)}  \, G_E^{-1 d b} \Big]
=  \kappa^2  \Big[ \theta_+^{a c}  { F}_{c d}^{(a)}  \theta_+^{d b}  - \theta_-^{a c}  {F}_{c d}^{(a)}  \theta_-^{d b}  \Big]       \, .
\end{eqnarray}
Note that neither of these depend on  the symmetric part ${F}_{a b}^{(s)}$. Because, according to Eq.~(\ref{eq:FcF0}) $ {\cal F}_{a b}^{(a)} = {F}_{a b}^{(a)}$, we can rewrite the above equations in terms of effective field strength   $ {\cal F}_{a b}^{(a)}$,
\begin{eqnarray}\label{eq:dFcFDic}
^\star {\cal F}^{a b}_{(a)} =   - \kappa^2 \theta^{a c}  {\cal F}_{c d}^{(a)}  \, \theta^{d b} - G_E^{-1 ac}  {\cal F}_{c d}^{(a)}  \, G_E^{-1 d b}
= - \frac{\kappa^2}{2} \Big[ \theta_+^{a c}  {\cal  F}_{c d}^{(a)}  \theta_+^{d b}  + \theta_-^{a c}  {\cal F}_{c d}^{(a)}  \theta_-^{d b}  \Big]       \, ,
\end{eqnarray}
and
\begin{eqnarray}\label{eq:dFcFDisc}
^\star {\cal F}^{a b}_{(s)} =   - 2 \kappa \Big[  G_E^{-1 a c}  {\cal F}_{c d}^{(a)}  \, \theta^{d b} +  \theta^{a c}   {\cal F}_{c d}^{(a)}  \, G_E^{-1 d b} \Big]
=  \kappa^2  \Big[ \theta_+^{a c}  {\cal  F}_{c d}^{(a)}  \theta_+^{d b}  - \theta_-^{a c}  {\cal F}_{c d}^{(a)}  \theta_-^{d b}  \Big]       \, .
\end{eqnarray}

In the Dirichlet sector of   T-dual theory we can reexpress the  term  $^\star S_A^D [y]$   in the ``standard" form where the effective vector fields  $^\star {\cal A}^a_\alpha$ are multiplied by the
${\dot y}_a^\alpha = \{  {\dot y}_a , - y_a^\prime \} = \eta^{\alpha \beta} \partial_\beta y_a$, so that the term with vector background fields takes the form
\begin{eqnarray}\label{eq:dcAdx}
^\star S_A^D [y] &=  2\kappa \eta^{\alpha \beta} \int d\tau \Big( {}^\star {\cal A}^a_\alpha [V] \, \partial_\beta {y}_a/_{\sigma=\pi}  - {}^\star {\cal A}^a_\alpha [V] \,\partial_\beta {y}_a/_{\sigma=0}     \Big) \,  \nonumber\\
& =  2\kappa \int d\tau \Big( {}^\star {\cal A}^a_\alpha [V] \, {\dot y}_a^\alpha/_{\sigma=\pi}  - {}^\star {\cal A}^a_\alpha [V] \, {\dot y}_a^\alpha/_{\sigma=0}     \Big) \,  ,
\end{eqnarray}
where ${}^\star {\cal A}^a_0 [V]$ and ${}^\star {\cal A}^a_1 [V]$ have been defined in Eq.~(\ref{eq:calAcd}).

\subsubsection{The case of Neumann vector fields}

For the T-dual  case,  corresponding to the Neumann vector field $ A^i_N (V)$,  we define the field strength
\begin{equation}\label{eq:Fabi}
{}^\star {\cal F}^{i j} = {}^\star {\cal F}^{i j}_{(a)} + \frac{1}{2} \, {}^\star {\cal F}^{i j}_{(s)} \, ,
\end{equation}
with the relation
\begin{eqnarray}\label{eq:dttots0}
& ^\star S_A^N [y] =  2 \kappa \int d \tau \Big[ \Big( {}^\star  A^i_N (V) {\dot y}_i  \Big)/_{\sigma=\pi} -
\Big( {}^\star  A^i_N (V) {\dot y}_i   \Big)/_{\sigma=0}  \Big]   \,   \nonumber\\
& =  \kappa \int d^2   \xi \, \partial_+  y_i \, {}^\star {\cal F}^{i j} \, \partial_- y_j   \,  .
\end{eqnarray}
After  some calculations we obtain
\begin{eqnarray}\label{eq:dFcF0}
^\star  {\cal F}^{i j}_{(a)} &=  - {}^\star B^{i k} \, ( {}^\star F^T)_k{}^j - {}^\star F^i {}_k  \, {}^\star B^{k j}  \,  ,
\end{eqnarray}
and
\begin{eqnarray}\label{eq:dFcF0s}
^\star  {\cal F}^{i j}_{(s)} &= -  {}^\star F^i {}_k  \, {}^\star G^{k j}  -  {}^\star G^{i k} \, ( {}^\star F^T)_k{}^j         \,  ,
\end{eqnarray}
where the coefficient ${}^\star F^i {}_j$ is defined as
\begin{eqnarray}\label{eq:sAN}
^\star A^i_N (V) = {}^\star A^i_0   - \frac{1}{2} \,  {}^\star F^i {}_j V^j    \, .
\end{eqnarray}

Using second equation in Eq.~(\ref{eq:Alc}),  Eq.~(\ref{eq:tdbfv}) and Eq.~(\ref{eq:sAN}),  we obtain ${}^\star F^i {}_j = G^{-1 ik} F_{k j}^{(s)}$ and consequently,
\begin{eqnarray}\label{eq:dcF}
 ^\star  {\cal F}^{i j}_{(a)} =
- \frac{\kappa}{4} \left(\theta^{i k} F_{k q}^{(s)} G^{-1 q j} + G^{-1 i k} F_{k q}^{(s)} \theta^{q j} \right)      \,  ,
\end{eqnarray}
and
\begin{eqnarray}\label{eq:dcFs}
 ^\star  {\cal F}^{i j}_{(s)} =  - \frac{1}{2} \left( G_E^{-1 i k} F_{k q}^{(s)} G^{-1 q j} + G^{-1 i k} F_{k q}^{(s)} G_E^{-1 q j} \right)  \,  .
\end{eqnarray}

We can eliminate $F_{i j}^{(a)}$ and $F_{i j}^{(s)}$  from Eqs.~(\ref{eq:FcF}),  (\ref{eq:FcFs})  and (\ref{eq:dcF})  and express $^\star  {\cal F}^{i j}_{(a)} $ in terms of  ${\cal F}_{i j}^{(a)} $ and ${\cal F}_{i j}^{(s)}$.
This is not a direct calculation, but we can check that expression
\begin{eqnarray}\label{eq:dFcFDi0}
^\star {\cal F}^{i j}_{(a)} =   - \kappa^2 \theta^{i k}  {\cal F}_{k q}^{(a)} \, \theta^{q j} - G_E^{-1 i k }  {\cal F}_{k q}^{(a)} \, G_E^{-1 q j}
- \frac{\kappa}{2} \left(G_E^{-1 i k }  {\cal F}_{k q}^{(s)} \, \theta^{ q j} + \theta^{ i k }  {\cal F}_{k q}^{(s)} \, G_E^{-1 q j} \right)       \, ,
\end{eqnarray}
is a proper solution. Similarly, we can eliminate the same variables  $F_{i j}^{(a)}$ and $F_{i j}^{(s)}$  from Eqs.~(\ref{eq:FcF}),  (\ref{eq:FcFs})  and (\ref{eq:dcFs})  and express $^\star  {\cal F}^{i j}_{(s)} $ in terms of  ${\cal F}_{i j}^{(a)} $ and ${\cal F}_{i j}^{(s)}$:
\begin{eqnarray}\label{eq:dFcFDi0s}
^\star {\cal F}^{i j}_{(s)} =   - \kappa^2 \theta^{i k}  {\cal F}_{k q}^{(s)} \, \theta^{q j} - G_E^{-1 i k }  {\cal F}_{k q}^{(s)} \, G_E^{-1 q j}
- 2 \kappa \left(G_E^{-1 i k }  {\cal F}_{k q}^{(a)} \, \theta^{ q j} + \theta^{ i k }  {\cal F}_{k q}^{(a)} \, G_E^{-1 q j} \right)       \, .
\end{eqnarray}

Similarly, in the Neumann sector of T-dual theory we can reexpress the  term  $^\star S_A^N [y]$   in the form
\begin{eqnarray}\label{eq:dcAdxn}
^\star S_A^N [y] &=  2\kappa \eta^{\alpha \beta} \int d\tau \Big( ^\star {\cal A}^i_\alpha [V] \, \partial_\beta {y}_i/_{\sigma=\pi}  - ^\star {\cal A}^i_\alpha [V] \,\partial_\beta {y}_i/_{\sigma=0}     \Big) \,  \nonumber\\
& =  2\kappa \int d\tau \Big( ^\star {\cal A}^i_\alpha [V] \, {\dot y}_i^\alpha/_{\sigma=\pi}  - ^\star {\cal A}^i_\alpha [V] \, {\dot y}_i^\alpha/_{\sigma=0}     \Big) \,  ,
\end{eqnarray}
where the effective vector fields  $^\star {\cal A}^i_\alpha [V] =\{ {}^\star {\cal A}^i_0 [V]\, , {}^\star {\cal A}^i_1 [V]   \} $ introduced in Eq.~(\ref{eq:calAcd}) are multiplied by ${\dot y}_i^\alpha = \{  {\dot y}_i , - y_i^\prime \} = \eta^{\alpha \beta} \partial_\beta y_i$.
We have put it in the suggestive form of Eq.~(\ref{eq:dcAdx})  although it is  much simpler. Because, according to
Eq.~(\ref{eq:calAcd}),   $ ^\star {\cal A}^i_1 [V]=0$, it has a form
\begin{eqnarray}\label{eq:dttots0c}
& ^\star S_A^N [y] =  2 \kappa \int d \tau \Big[ \Big( {}^\star  {\cal A}^i_0 (V) {\dot y}_i  \Big)/_{\sigma=\pi} -
\Big( {}^\star  {\cal A}^i_0 (V) {\dot y}_i   \Big)/_{\sigma=0}  \Big]   \,    .
\end{eqnarray}

\subsection{ T-dual field strength in terms of initial one}

Let us introduce  the complete field strengths
\begin{equation}\label{eq:Fmn}
{\cal F}_{\mu \nu} = {\cal F}_{\mu \nu}^{(a)} + \frac{1}{2} {\cal F}_{\mu \nu}^{(s)} \, ,  \quad
{\cal F}_{\mu \nu}^{(a)} = \left (
\begin{array}{cc}
{\cal F}_{a b}^{(a)} &  0 \\
0  &  {\cal F}_{i j}^{(a)}
\end{array}\right )\,  , \quad
{\cal F}_{\mu \nu}^{(s)} = \left (
\begin{array}{cc}
0 &  0 \\
0  &  {\cal F}_{i j}^{(s)}
\end{array}\right )\,  ,
\end{equation}
which contain the Neumann parts  ${\cal F}_{a b}^{(a)}$ defined in Eq.~(\ref{eq:FcF0}), as well as the Dirichlet ones ${\cal F}_{i j}^{(a)}$ and ${\cal F}_{i j}^{(s)}$ defined in Eqs.~(\ref{eq:FcF}) and (\ref{eq:FcFs}).
Then, taking into account that  according to Eq.~(\ref{eq:cvf})  ${\cal A}_{1 a} (x) =0$,    we can rewrite the action with vector background fields as:

\begin{eqnarray}\label{eq:vbf}
&S_A (x) =S_A^N (x) + S_A^D (x)
= 2\kappa \eta^{\alpha \beta} \int d\tau \Big(  {\cal A}_{\alpha \mu} [x] \, \partial_\beta x^\mu/_{\sigma=\pi}  -  {\cal A}_{\alpha \mu} [x] \,\partial_\beta x^\mu/_{\sigma=0}     \Big) \nonumber\\
&= \kappa \int d^2 \xi \partial_+ x^\mu   {\cal F}_{\mu \nu}   \partial_-  x^\nu \,   ,
\end{eqnarray}
where the expressions for the terms  $S_A^N (x)$ and  $S_A^D (x)$  have been defined in  Eqs.~(\ref{eq:cAdx})  and  (\ref{eq:cAdxd}).
Note that $S_A (x)$   has the same form as the initial action and according to  Eq.~(\ref{eq:Fmn})  contains both symmetric and antisymmetric parts.
So, the other way to introduce vector background fields is to substitute the Kalb-Ramond field $B_{\mu \nu}$ and metric  $G_{\mu \nu}$   with
\begin{eqnarray}
B_{\mu \nu}  \to {\cal B}_{\mu \nu} = B_{\mu \nu} + {\cal F}_{\mu \nu}^{(a)} \, ,   \qquad
G_{\mu \nu}  \to {\cal G}_{\mu \nu} = G_{\mu \nu} + {\cal F}_{\mu \nu}^{(s)}  \, .
\end{eqnarray}
Note that according to Eq.~(\ref{eq:phv}) the new variables are just gauge invariant ones.  Then the open string action takes the form
\begin{eqnarray}\label{eq:Aop}
S_{open} = \kappa \int d^2 \xi \partial_+ x^\mu \Big(  {\cal B}_{\mu \nu}  +  \frac{1}{2} {\cal G}_{\mu \nu} \Big) \partial_-  x^\nu \, .
\end{eqnarray}

Because  all background fields in this  action are constant, we already know the form of T-dual fields for such an action. In analogy with  Eq.~(\ref{eq:tdbf}) we have
\begin{equation}\label{eq:tdbfd}
{}^\star {\cal G}^{\mu\nu} =
({\cal G}_{E}^{-1})^{\mu\nu}, \quad
^\star {\cal B}^{\mu\nu} =
\frac{\kappa}{2}
{\Theta}^{\mu\nu}  \, ,
\end{equation}
where according to Eq.~(\ref{eq:GEt}),
\begin{eqnarray}
{\cal G}^E_{\mu\nu} & \equiv  {\cal G}_{\mu\nu}-4({\cal B} {\cal G}^{-1}{\cal B})_{\mu\nu}, \qquad
\Theta^{\mu\nu} & \equiv  -\frac{2}{\kappa} ({\cal G}^{-1}_{E}{\cal B} {\cal G}^{-1})^{\mu\nu} \, .
\end{eqnarray}

Taking into account that the vector background fields  and consequently their field strengths  are infinitesimal, we can separate the infinitesimal part of $^\star {\cal B}^{\mu\nu}$,
\begin{eqnarray}\label{eq:FTd}
{}^\star {\cal  F}^{\mu \nu}_{(a)} &=
-  G^{-1 \mu \rho}_E  {\cal  F}_{\rho \sigma}^{(a)}  G^{-1 \sigma \nu}_E  - \kappa^2 \theta^{\mu \rho} {\cal F}_{\rho \sigma}^{(a)} \theta^{\sigma \nu}  -
\frac{\kappa}{2}   \left(  G^{-1 \mu \rho}_E  {\cal  F}_{\rho \sigma}^{(s)}  \theta^{\sigma \nu}    +
\theta^{\mu \rho} {\cal F}_{\rho \sigma}^{(s)}  G^{-1 \sigma \nu}_E  \right)   \nonumber\\
&= - \frac{\kappa^2}{2} \left( \theta_+^{\mu \rho}  {\cal  F}_{\rho \sigma}^{(a)} \theta_+^{\sigma \nu}  + \theta_-^{\mu \rho}  {\cal F}_{\rho \sigma}^{(a)} \theta_-^{\sigma \nu}  \right)
+ \frac{\kappa^2}{4} \left( \theta_+^{\mu \rho}  {\cal  F}_{\rho \sigma}^{(s)} \theta_+^{\sigma \nu}  - \theta_-^{\mu \rho}  {\cal F}_{\rho \sigma}^{(s)} \theta_-^{\sigma \nu}  \right)   \, .
\end{eqnarray}
and the infinitesimal part of ${}^\star {\cal G}^{\mu\nu}$,
\begin{eqnarray}\label{eq:FTds}
{}^\star {\cal  F}^{\mu \nu}_{(s)} &=
-  G^{-1 \mu \rho}_E  {\cal  F}_{\rho \sigma}^{(s)}  G^{-1 \sigma \nu}_E  - \kappa^2 \theta^{\mu \rho} {\cal F}_{\rho \sigma}^{(s)} \theta^{\sigma \nu}
- 2 \kappa    \left(  G^{-1 \mu \rho}_E  {\cal  F}_{\rho \sigma}^{(a)}  \theta^{\sigma \nu}    +
\theta^{\mu \rho} {\cal F}_{\rho \sigma}^{(a)}  G^{-1 \sigma \nu}_E  \right)   \nonumber\\
&= - \frac{\kappa^2}{2} \left( \theta_+^{\mu \rho}  {\cal  F}_{\rho \sigma}^{(s)} \theta_+^{\sigma \nu}  + \theta_-^{\mu \rho}  {\cal F}_{\rho \sigma}^{(s)} \theta_-^{\sigma \nu}  \right)
+ \kappa^2  \left( \theta_+^{\mu \rho}  {\cal  F}_{\rho \sigma}^{(a)} \theta_+^{\sigma \nu}  - \theta_-^{\mu \rho}  {\cal F}_{\rho \sigma}^{(a)} \theta_-^{\sigma \nu}  \right)   \, .
\end{eqnarray}

Here $\theta_\pm^{\mu\nu}$ has been defined in Eq.~(\ref{eq:tpm}) and  the complete T-dual field strengths
\begin{equation}\label{eq:Fmnd}
{}^\star {\cal F}^{\mu \nu} = {}^\star {\cal F}^{\mu \nu}_{(a)} + \frac{1}{2} {}^\star {\cal F}^{\mu \nu}_{(s)} \, ,  \quad
{}^\star {\cal F}^{\mu \nu}_{(a)} = \left (
\begin{array}{cc}
{}^\star {\cal F}^{a b}_{(a)} &  0 \\
0  &  {}^\star {\cal F}^{i j}_{(a)}
\end{array}\right )\,  , \quad
{}^\star {\cal F}^{\mu \nu}_{(s)} = \left (
\begin{array}{cc}
{}^\star {\cal F}^{a b}_{(s)} &  0 \\
0  &  {}^\star {\cal F}^{i j}_{(s)}
\end{array}\right )\,  ,
\end{equation}
contain Neumann parts    ${}^\star {\cal F}^{a b}_{(a)}$ and  ${}^\star {\cal F}^{a b}_{(s)}$  as well as  Dirichlet ones   ${}^\star {\cal F}^{i j}_{(a)}$ and
${}^\star {\cal F}^{i j}_{(s)}$.
Therefore, taking into account that ${\cal F}_{a b}^{(s)}=0$,  Eq.~(\ref{eq:FTd}) is in complete agreement with Eq.~(\ref{eq:dFcFDic})  and  Eq.~(\ref{eq:dFcFDi0}), as well as Eq.(\ref{eq:FTds}) being in complete agreement with  Eqs.~(\ref{eq:dFcFDisc})   and (\ref{eq:dFcFDi0s}).

%%%%%%%%%%%%%%%%%%%%%%%%%%%%%%%%%%%%%%%%%%%%%%%%%%%%%%%%%%%%%%%%%%%%%%%%%%%%%%%%%%%%%
\cleq

\section{ Field strength for non-geometric theories}

In our approach, a characteristic feature of non-geometric theories are background dependence on variable $V^\mu$, which includes dependence on
both T-dual coordinate $y_\mu$ and its double $\tilde y_\mu$.
In Refs.~\cite{DNS} and \cite{DNS2} it was shown that $V^\mu$-dependence produces non-commutativity  and non-associativity  of the closed string coordinates. It is also the origin of difficulties in defining field strength for non-geometric theories. In this section we offer a solution to this problem and define the field strength of non-geometric theories which include derivation with respect to
both $y_\mu$  and $\tilde y_\mu$.

In geometric theories the field strength for an Abelian vector field ${\cal A}_\mu (x)$ is defined simply as
${\cal F}_{\mu \nu}= \partial_\mu {\cal A}_\nu - \partial_\nu {\cal A}_\mu $, where with $\partial_\mu$ we  denote  derivation with respect to the variable $x^\mu$. In non-geometric theories the vector field ${\cal A}_\mu (V)$ depends on $V^\mu = - \kappa \, \theta^{\mu \nu} y_\nu + G^{-1 \mu \nu}_E \, {\tilde y}_\nu$, which means that it depends on two variables $y_\mu$ and ${\tilde y}_\mu$. Generally speaking, in order to obtain  the field strength for non-geometric theories, we should have the derivative  with respect to both variables $y_\mu$ and ${\tilde y}_\mu$.

\subsection{ Non-geometric field strengths in terms of effective gauge fields}

For the initial  theory, according to Eqs.~(\ref{eq:FcF0}), (\ref{eq:FcF}) and (\ref{eq:FcFs}) and taking into account that ${\cal A}_{1 a} = 0$,   we have
\begin{eqnarray}\label{eq:FcFsmu}
{\cal F}_{\mu \nu}^{(a)} = \partial_\mu {\cal A}_{0 \nu} (x) - \partial_\nu {\cal A}_{0 \mu} (x) \,  , \qquad
{\cal F}_{\mu \nu}^{(s)} =  2 \Big[  \partial_\mu {\cal A}_{1 \nu} (x) + \partial_\nu {\cal A}_{1 \mu} (x) \Big] \,   .
\end{eqnarray}
The antisymmetric part has a standard form, but we also obtain a non-trivial symmetric part.

Let us consider the field strengths of T-dual  non-geometric theories.
Until now we obtained the complete expressions for T-dual field strengths  ${}^\star  {\cal F}^{\mu \nu}$ for non-geometric theories. The next step is to write out these
expressions in terms of  derivatives   of T-dual gauge fields ${}^\star {\cal A}^a_0 (V)$  and ${}^\star {\cal A}^a_1 (V)$   with respect to  variables $y_\mu$ and ${\tilde y}_\mu$, and find local gauge symmetries in such cases.

With the help of the first line in  Eq.~(\ref{eq:calAcd}) and Eq.~(\ref{eq:solV})   we find the antisymmetric parts,
\begin{eqnarray}\label{eq:Fabn0y}
& {}^\star_{y}  {\cal F}_{0 (a)}^{a b} \equiv   \partial_y^a \, \, {}^\star {\cal A}^b_0 (V)  - \partial_y^b \, \, {}^\star {\cal A}^a_0 (V) = - \kappa^2 (\theta F_{(a)} \theta)^{a b} \,  , \nonumber\\
& {}^\star_{\tilde y}  {\cal F}_{1 (a)}^{a b} \equiv \partial_{\tilde y}^a \, \, {}^\star {\cal A}^b_1 (V) - \partial_{\tilde y}^b \, \, {}^\star {\cal A}^a_1 (V) =
 - (G^{-1}_E  F_{(a)} G^{-1}_E)^{a b} \,  ,
\end{eqnarray}
and symmetric parts,
\begin{eqnarray}\label{eq:Fab0ty}
&  {}^\star_{y}  {\cal F}_{1 (s)}^{a b} \equiv \partial_{y}^a \, \, {}^\star {\cal A}^b_1 (V) + \partial_{ y}^b \, \, {}^\star {\cal A}^a_1 (V) =  - \frac{\kappa}{2}(G^{-1}_E  F_{(a)} \theta + \theta  F_{(a)} G^{-1}_E )^{a b}  \, , \nonumber\\
& {}^\star_{\tilde y}  {\cal F}_{0 (s)}^{a b} \equiv \partial_{\tilde y}^a \, \, {}^\star {\cal A}^b_0 (V) + \partial_{\tilde y}^b \, \, {}^\star {\cal A}^a_0 (V) =  - \frac{\kappa}{2}(G^{-1}_E  F_{(a)} \theta + \theta  F_{(a)} G^{-1}_E )^{a b} \, ,
\end{eqnarray}
where with $\partial^a_y$ and $\partial^a_{\tilde y}$ we denote partial derivations with respect to $y_a$ and ${\tilde y}_a$.

With the help of the last line in Eq.~(\ref{eq:calAcd}) and  Eq.~(\ref{eq:solV}),   for the $i j$ sector,  we have
\begin{eqnarray}\label{eq:Fij0y}
& {}^\star_{y}  {\cal F}_{0 (a)}^{i j} \equiv   \partial_y^i \, \, {}^\star {\cal A}^j_0 (V)  - \partial_y^j \, \, {}^\star {\cal A}^i_0 (V) =
- \frac{\kappa}{4}(G^{-1}  F_{(s)} \theta + \theta  F_{(s)} G^{-1} )^{i j}     \,  , \nonumber\\
& {}^\star_{\tilde y}  {\cal F}_{0  {(s)}}^{i j} \equiv   \partial_{\tilde y}^i \, \, {}^\star {\cal A}^j_0 (V)  + \partial_{\tilde y}^j \, \, {}^\star {\cal A}^i_0 (V) =
- \frac{1}{4}(G^{-1}  F_{(s)} G^{-1}_E  + G^{-1}_E   F_{(s)} G^{-1} )^{i j}  \,  .
\end{eqnarray}

Because according to  Eq.~(\ref{eq:calAcd}),   ${}^\star {\cal A}^i_1 (V) = 0$,  all corresponding field strengths (both symmetric and antisymmetric parts) vanish, ${}^\star_{y}  {\cal F}_{1}^{i j} = 0 = {}^\star_{\tilde y}  {\cal F}_{1}^{i j}$.

Comparing Eq.~(\ref{eq:Fabn0y})    with Eq.~(\ref{eq:dFcFDi}) we find
\begin{eqnarray}\label{eq:TDfa}
^\star  {\cal F}^{a b}_{(a)}  = \, ^\star_{y}  {\cal F}_{0 (a)}^{a b} +\,  ^\star_{\tilde y}  {\cal F}_{1 (a)}^{a b} = \partial_y^a \, \, {}^\star {\cal A}^b_0 (V) - \partial_y^b \, \, {}^\star {\cal A}^a_0 (V) +  \partial_{\tilde y}^a \, \,
{}^\star {\cal A}^b_1 (V) - \partial_{\tilde y}^b \, \, {}^\star {\cal A}^a_1 (V) \, .
\end{eqnarray}
Similarly, comparing  Eq.~(\ref{eq:Fab0ty})    with Eq.~(\ref{eq:dFcFDis}) we have
\begin{eqnarray}\label{eq:TDfs}
^\star  {\cal F}^{a b}_{(s)}  = 2 \left( {}^\star_{\tilde y}  {\cal F}_{0 (s)}^{a b} +\,  ^\star_{y}  {\cal F}_{1 (s)}^{a b} \right)  =
2  \left( \partial_{\tilde y}^a \, \, {}^\star {\cal A}^b_0 (V) + \partial_{\tilde y}^b \, \, {}^\star {\cal A}^a_0 (V) +  \partial_{y}^a    \, \,
{}^\star {\cal A}^b_1 (V) + \partial_{y}^b \, \, {}^\star {\cal A}^a_1 (V) \right)   \, .
\end{eqnarray}

For the Dirichlet sector, comparing  Eq.~(\ref{eq:dcF})    with the first relation in Eq.~(\ref{eq:Fij0y}), we obtain
\begin{eqnarray}\label{eq:TDfai}
^\star  {\cal F}^{i j}_{(a)}  = \, ^\star_{y}  {\cal F}_{0 (a)}^{i j}  = \partial_y^i \, \, {}^\star {\cal A}^j_0 (V) - \partial_y^j \, \, {}^\star {\cal A}^i_0 (V)  \, ,
\end{eqnarray}
while comparing  Eq.~(\ref{eq:dcFs})    with the second relation in Eq.~(\ref{eq:Fij0y}) we have
\begin{eqnarray}\label{eq:TDfais}
^\star  {\cal F}^{i j}_{(s)}  = 2 \, \, {}^\star_{\tilde y}  {\cal F}_{0 (s)}^{i j}  = 2 \Big( \partial_{\tilde y}^i \, \, {}^\star {\cal A}^j_0 (V) + \partial_{\tilde y}^j \, \, {}^\star {\cal A}^i_0 (V) \Big) \, .
\end{eqnarray}

Taking  into account that ${}^\star {\cal A}^i_1 (V) = 0$,  we can conclude that the same relations are valid for both Neumann and Dirichlet  sectors. Consequently, such a form is valid for complete field strengths, and
with $\mu, \nu$  indices  we have
\begin{eqnarray}\label{eq:TDfa}
^\star  {\cal F}^{\mu \nu}_{(a)}  =
 \partial_y^\mu \, \, {}^\star {\cal A}^\nu_0 (V) - \partial_y^\nu \, \, {}^\star {\cal A}^\mu_0 (V) +  \partial_{\tilde y}^\mu \, \, {}^\star {\cal A}^\nu_1 (V) - \partial_{\tilde y}^\nu \, \, {}^\star {\cal A}^\mu_1 (V)  \,  , \nonumber\\  {}^\star  {\cal F}^{\mu \nu}_{(s)}  =
2  \left[ \partial_{\tilde y}^\mu \, \, {}^\star {\cal A}^\nu_0 (V) + \partial_{\tilde y}^\nu \, \, {}^\star {\cal A}^\mu_0 (V) +  \partial_{y}^\mu    \, \,
{}^\star {\cal A}^\nu_1 (V) + \partial_{y}^\nu \, \, {}^\star {\cal A}^\mu_1 (V)  \right]  \, .
\end{eqnarray}

If we define $y_\mu^\alpha = \{  y_\mu^0 = y_\mu , \,  y_\mu^1 = -{\tilde y}_\mu \} $  and
$\partial_\alpha^\mu  \equiv \frac{\partial}{\partial y^\alpha_\mu} = \{\frac{\partial}{\partial y_\mu} \,  , \frac{\partial}{\partial {\tilde y}_\mu}    \}  $,  we can rewrite the above equations in a compact form,
\begin{eqnarray}\label{eq:sFa}
{}^\star {\cal F}^{\mu \nu}_{(a)} =  \eta^{\alpha \beta} \Big[  \partial_\alpha^\mu \, \, {}^\star {\cal A}^\nu_\beta  (V) - \partial_\alpha^\nu \, \, {}^\star {\cal A}^\mu_\beta (V) \Big]  \,  , \quad
{}^\star {\cal F}^{\mu \nu}_{(s)} = -2  \varepsilon^{\alpha \beta} \Big[  \partial_\alpha^\mu \, \, {}^\star {\cal A}^\nu_\beta  (V) + \partial_\alpha^\nu \, \, {}^\star {\cal A}^\mu_\beta (V) \Big]  \,  .
\end{eqnarray}
Finally, we have
\begin{eqnarray}\label{eq:cfsF}
{}^\star {\cal F}^{\mu \nu} =& {}^\star {\cal F}^{\mu \nu}_{(a)} + \frac{1}{2} {}^\star {\cal F}^{\mu \nu}_{(s)}
=  \eta^{\alpha \beta} \Big[  \partial_\alpha^\mu \, \, {}^\star {\cal A}^\nu_\beta  (V) - \partial_\alpha^\nu \, \, {}^\star {\cal A}^\mu_\beta (V) \Big]
-  \varepsilon^{\alpha \beta} \Big[  \partial_\alpha^\mu \, \, {}^\star {\cal A}^\nu_\beta  (V) + \partial_\alpha^\nu \, \, {}^\star {\cal A}^\mu_\beta (V) \Big] \nonumber\\
=& (\eta^{\alpha \beta} - \varepsilon^{\alpha \beta} ) \, \partial_\alpha^\mu \, \, {}^\star {\cal A}^\nu_\beta  (V) - (\eta^{\alpha \beta} +
\varepsilon^{\alpha \beta} ) \, \partial_\alpha^\nu \, \, {}^\star {\cal A}^\mu_\beta (V)  \,  .
\end{eqnarray}

We can check this expression in another way. From Eqs.~(\ref{eq:dcAdx}) and (\ref{eq:dcAdxn}) we have,
\begin{eqnarray}\label{eq:sSay}
{}^\star S_A [y] = {}^\star S_A^D [y] + {}^\star S_A^N [y] =
2\kappa \eta^{\alpha \beta} \int d\tau \Big( ^\star {\cal A}^\mu_\alpha [V] \, \partial_\beta {y}_\mu/_{\sigma=\pi}  - ^\star {\cal A}^\mu_\alpha [V] \,\partial_\beta {y}_\mu/_{\sigma=0}     \Big) \,  \,  .
\end{eqnarray}
After transition from integration over $\tau$ to integration over $d^2 \xi = d\tau d \sigma$ and  partial integration over $\tau$,  we obtain
\begin{eqnarray}\label{eq:sScF}
{}^\star S_A [y] = \kappa \int d^2 \xi \partial_+ y_\mu  {}^\star {\cal F}^{\mu \nu}   \partial_-  y_\nu \,   ,
\end{eqnarray}
where ${}^\star {\cal F}^{\mu \nu}$ is just Eq.~(\ref{eq:cfsF}), obtained previously in  the other way.

Let us stress that the field strength of the initial  theory is a particular case of Eq.~(\ref{eq:cfsF}).  In fact, in that case background fields depend only on $x^\mu$ and not on ${\tilde x}^\mu$.
So, if in Eq.(\ref{eq:cfsF}) we omit terms which contain derivatives with respect to the tilde variable ${\tilde y}_\mu= - y^1_\mu$, we obtain a relation of the same form as that in Eq.~(\ref{eq:FcFsmu}).

Equation~(\ref{eq:cfsF}) we can consider as a general definition of the field strength for both geometric and  non-geometric theories. Note that besides the antisymmetric part  ${}^\star {\cal F}^{\mu \nu}_{(a)} $  it also contains a symmetric part ${}^\star {\cal F}^{\mu \nu}_{(s)}$. In the definition of both parts,  the  derivatives with respect to both  T-dual coordinate $y_\mu$ and to its double  ${\tilde y}_\mu$  contribute.

The unusual form of $^\star  {\cal F}^{\mu \nu}$ is a consequence of two facts. First, the T-dual vector field   $^\star {A}^a_D (V)$  is not multiplied by  ${\dot y}_a$
but by T-dual  $\sigma$-momentum   $^\star G^{-1}_{a b} {}^\star \gamma^b_{(0)}$; and second,
the T-dual vector fields depend on $V^\mu$   (see Eq.~(\ref{eq:solV}))  which is a function of both  $y_\mu$ and ${\tilde y}_\mu$.

%%%%%%%%%%%%%%%%%%%%%%%%%%%%%%%%%%%%%%%%%%%%%%%%%%%%%%%%%%%%%%%%%%%%%

\cleq

\section{Genuinely non-geometric theories}

Until now we have used a generalized Buscher procedure to establish a new structure for non-geometric theories defined in terms of effective vector fields ${\cal A}^\mu_0 (V),  {\cal A}^\mu_1 (V)$ and effective field strength
${\cal F}^{\mu \nu}$. It is important to stress that effective vector fields are not independent. The initial vector fields  are connected with Eqs.~(\ref{eq:cvf}) and (\ref{eq:cvfr}), and the T-dual ones with
Eq.~(\ref{eq:cvfd}).

Now we are able  to separate from the Buscher approach and establish new kinds of non-geometric theories.  We can preserve the  obtained structure Eqs.~(\ref{eq:cfsF})-(\ref{eq:sScF}),  and omit the relation between effective vector fields.
Consequently, in all obtained theories we will have a nontrivial   field ${\cal A}_{1 a} (V)$.  Moreover, we can define new background field dependence on the arguments. As well as T-dual background fields dependent on
\begin{eqnarray}\label{eq:Vmy}
V^\mu (y)=  - \kappa \theta^{\mu \nu} y_\nu + G_E^{-1 \mu \nu} {\tilde y}_\nu  = -2 \, {}^\star B^{\mu \nu} y_\nu + \, {}^\star  G^{\mu \nu} {\tilde y}_\nu \,   ,
\end{eqnarray}
(see Eq.~(\ref{eq:solV})), which is a solution for $x^\mu$ of the  finite part (for ${\cal A}_{\pm \mu} = 0$) of  T-dual transformation laws (\ref{eq:tdtr}), we will take that  the initial background fields are dependent on
$V_\mu (x)$, which is a solution for $y_\mu$ of the finite part of  the inverse T-dual transformation laws (\ref{eq:itdtr}). So it takes the form
\begin{eqnarray}\label{eq:Vmx}
V_\mu (x)= -2 B_{\mu \nu} x^\nu +  G_{\mu \nu} {\tilde x}^\nu \,   ,
\end{eqnarray}
and depends on ${\tilde x}^\mu = \int (d \tau x^{\prime \mu}  + d \sigma {\dot x}^\mu) $, which makes the theory non-geometric. Therefore all theories, including the initial one,  will be non-geometric.
Our vector fields of genuinely non-geometric theories are ${\cal A}_{\pm \mu} [V_\mu (x)]$ and
 ${}^\star {\cal A}^\mu_\pm [V^\mu (y)]$, where the arguments are defined in the above expressions.

Now new duality transformations take a simple form,
\begin{eqnarray}\label{eq:gtd}
{\cal A}_{\pm \mu} [V_\mu (x)]  \to    {}^\star {\cal A}^\mu_\pm [V^\mu (y)] =  \kappa \, \theta_\pm^{\mu \nu}   {\cal A}_{\pm \nu} [V^\mu (y)]  \,  .
\end{eqnarray}
Then, for example, the inverse T-dual transformation produces non-trivial relations
\begin{eqnarray}\label{eq:ina1}
  2 B_{\mu \nu} \, {}^\star {\cal A}^\nu_0 [V^\mu (y)] - G_{\mu \nu} \, {}^\star {\cal A}^\nu_1 [V^\mu (y)]  \to  {\cal A}_{0 \mu} [V_\mu (x)]  \,  ,  \nonumber \\
 2 B_{\mu \nu} \, {}^\star {\cal A}^\nu_1 [V^\mu (y)] - G_{\mu \nu} \, {}^\star {\cal A}^\nu_0 [V^\mu (y)]  \to {\cal A}_{1 \mu} [V_\mu (x)]  \, .
\end{eqnarray}
The constraints (\ref{eq:cvfd}) on the T-dual effective fields force the Neumann part to zero, ${\cal A}_{1 a}=0$, but without
these constraints  ${\cal A}_{1 a}$ is non-trivial.  Also, the fields ${\cal A}_{\alpha \mu}$ depend on $V_\mu (x)$
and we have truly non-geometric theories. In Section 6.2.  we will  introduce a matter non-geometric field.
The part $\psi_1 (V)$, corresponding to gauge field ${\cal A}^\mu_1 (V)$, prevents regression to the geometric theory after T-dualization.

There exist a few different approaches to genuinely non-geometric theories. For more details see  Ref.~\cite{CJL} and references therein.

\subsection{Local gauge symmetries of non-geometric theories}

 We are ready to find  gauge transformations of the vector fields in non-geometric theories.
 To be definite, we will use T-dual fields  $^\star {\cal A}^\mu_\alpha [V^\mu (y)]$, but similar expressions are valid for initial fields  ${\cal A}_{\alpha \mu} [ V_\mu (x)]$.
 By our definition the action (\ref{eq:vbf}) is proportional to field strength, and consequently it is gauge invariant. So,  it is enough to find a transformation which leaves the action (\ref{eq:vbf}) unchanged. It is easy to see that transformation,
\begin{eqnarray}\label{eq:nggt}
 ^\star {\cal A}^\mu_\alpha [V]     \to  \, \,   ^\star {\cal A}^\mu_\alpha [V]  + \partial_\alpha^\mu \lambda_\alpha (y^\alpha)   \, , \qquad  ( \lambda_\alpha (y^\alpha) \equiv \{ \lambda_0 (y)\, , \lambda_1 (\tilde y)  \} )
\end{eqnarray}
or equivalently in components,
\begin{eqnarray}\label{eq:nggtc}
 ^\star {\cal A}^\mu_0 [V]     \to  \, \,   ^\star {\cal A}^\mu_0 [V]  + \partial_y^\mu \lambda_0 (y)   \, , \qquad
 ^\star {\cal A}^\mu_1 [V]     \to  \, \,   ^\star {\cal A}^\mu_1 [V]  + \partial_{\tilde y}^\mu \lambda_1 ({\tilde y})   \, ,
\end{eqnarray}
satisfy this condition because
\begin{eqnarray}\label{eq:pnggt}
\eta^{\alpha \beta}  \int d\tau  \, \partial_\alpha^\mu  \lambda_\alpha (y^\alpha)  \, \,\partial_\beta  { y}_\mu/_{\partial\Sigma}
= \int d\tau  \,\Big( \partial_y^\mu  \lambda_0 (y) \, {\dot y}_\mu  - \partial_{\tilde y}^\mu  \lambda_1 ({\tilde y}) \, {\dot{\tilde  y}}_\mu \Big)/_{\partial\Sigma}   \nonumber\\
= \int d\tau  \Big(  {\dot \lambda}_0  -  {\dot \lambda}_1   \Big)/_{\partial\Sigma} =0     \, .
\end{eqnarray}

Consequently, the expression for field strength in  Eq.~(\ref{eq:cfsF}) should be invariant under gauge transformations,
\begin{eqnarray}\label{eq:nggti}
\delta \, \,  ^\star {\cal A}^\mu_\alpha [V]   =    \partial_\alpha^\mu \lambda_\alpha (y^\alpha)    \, ,
\end{eqnarray}
or in components,
\begin{eqnarray}\label{eq:dnggtc}
\delta \, \, ^\star {\cal A}^\mu_0 [V]   =   \partial_y^\mu \lambda_0 (y)   \, , \qquad
\delta \, \, ^\star {\cal A}^\mu_1 [V]   =    \partial_{\tilde y}^\mu \lambda_1 ({\tilde y})   \, .
\end{eqnarray}
It easy to check that this is true. In fact variation of the antisymmetric part (the coefficient in front of $\eta^{\alpha \beta}$) vanishes  in the same way as in geometric
theory (partial derivatives commute). Variation of the symmetric part (the coefficient in front of $\varepsilon^{\alpha \beta}$) vanishes because we have derivatives with respect to  both $y_\mu$ and ${\tilde y}_\mu$ of the parameter $\lambda$ which depend on
only one of these variables.

The  transformation (\ref{eq:nggtc})  we can take as the definition of gauge transformations for non-geometric theories.

\subsection{Non-geometric matter fields}

In the description of T-dual non-geometric fields we introduced a pair of T-dual coordinates $y_\mu^0 =  y_\mu$ and $y_\mu^1 =  {\tilde y}_\mu$, as well as  a pair of effective vector fields ${\cal A}^\mu_0 (V)$ and  ${\cal A}^\mu_1 (V)$.
Each vector field transforms with its gauge parameter $\lambda_0(y)$ and $\lambda_1 ({\tilde y})$. So, it is natural to introduce a pair of spinor  matter fields $\psi_0 (V)$ and $\psi_1 (V)$ with Lagrangian
\begin{eqnarray}\label{eq:Lag}
{\cal L}  = {\bar \psi}_0 (V) i\gamma_\mu \partial^\mu_y  \psi_0 (V) + {\bar \psi}_1 (V) i\gamma_\mu \partial^\mu_{\tilde y}  \psi_1 (V)   \, .
\end{eqnarray}

As well as in the standard electrodynamics, it is invariant under  global symmetries,
\begin{eqnarray}\label{eq:ggt}
 \psi_0^{(\lambda_0)}  (V) = e^{-i \lambda_0}   \psi_0 (V)   \, , \qquad        \psi_1^{(\lambda_1)}  (V) = e^{-i \lambda_1}   \psi_1 (V)   \, .   \quad (\lambda_0, \lambda_1 = const)
\end{eqnarray}
Now, we can gauge these symmetries  requiring that Lagrangian (\ref{eq:Lag}) is invariant under corresponding local symmetries  with parameters $\lambda_0(y)$ and $\lambda_1 (\tilde y)$. This can be achieved by
introducing covariant derivatives,
\begin{eqnarray}\label{eq:cD}
\partial^\mu_y  \to D^\mu_y  = \partial^\mu_y  + i  \, \, ^\star {\cal A}^\mu_0  (V)  \, , \qquad   \partial^\mu_{\tilde y}   \to {\tilde D}^\mu_{\tilde y}   = \partial^\mu_{\tilde y}   + i \,  \, ^\star {\cal A}^\mu_1  (V)   \, .
\end{eqnarray}

With the help of Eqs.~(\ref{eq:nggtc}) and (\ref{eq:ggt}) we  can easily check that the covariant derivatives really transform as
\begin{eqnarray}\label{eq:cDtr}
 \left[ D^\mu_y  \psi_0 (V) \right]^{(\lambda_0)}  =  e^{-i \lambda_0(y)}   D^\mu_y  \psi_0 (V)    \, , \qquad   \left[ {\tilde D}^\mu_{\tilde y}  \psi_1 (V) \right]^{(\lambda_1)}
 =  e^{-i \lambda_1 (\tilde y)}   {\tilde D}^\mu_{\tilde y}  \psi_1 (V)      \, .
\end{eqnarray}

Consequently, the interaction Lagrangian obtains the form
\begin{eqnarray}\label{eq:Lagi}
{\cal L}_{int}  = - {\bar \psi}_0 (V) \gamma_\mu  \psi_0 (V) \, ^\star {\cal A}^\mu_0  (V) - {\bar \psi}_1 (V) \gamma_\mu   \psi_1 (V)  \, ^\star {\cal A}^\mu_1 (V)  \, .
\end{eqnarray}

It is possible to form the Lagrangian  for gauge fields in non-geometric theories by constructing the scalar from the gauge invariant field strength (\ref{eq:cfsF}). In analogy with electrodynamics we can write
${}^\star {\cal L} \sim   \, {}^\star  {\cal F}^{\mu \nu}  \, {}^\star  {\cal F}_{\mu \nu}$,  while in analogy with Born-Infeld theory we have
${}^\star {\cal L} \sim \sqrt{- \det(\eta_{\mu \nu} + 2 \pi \alpha^\prime {}^\star  {\cal F}_{\mu \nu} )}   $. The equations of motion and other features of non-geometric theories, which follow from these Lagrangians, will be discussed elsewhere.

We offer a possible interpretation of such a non-geometric theory. The fields  with index $0$,  $\psi_0 (V)$  and $^\star {\cal A}^\mu_0 (V)$, are standard ones and we can suppose that they
represent  known spinor and gauge fields. The fields  with index $1$, $\psi_1 (V)$  and $^\star {\cal A}^\mu_1 (V)$, are new ones and we can suppose that they  describe    some so far unknown physics.
It  might be interesting to consider its possible   relation with  dark matter and dark energy.

%%%%%%%%%%%%%%%%%%%%%%%%%%%%%%%%%%%%%%%%%%%%%%%%%%%%%%%%%%%%%%%%%%%%%

\cleq

\section{Example: three-torus with $D_1$-brane}

In this section we will take the example of a three-torus with $D_1$-brane.
%considering  in of the Refs.\cite{ALLP,ALLP1,HLZ,BL,DNS2}.
We will perform T-dualization along all coordinates and obtain the T-dual three-torus with $D_0$-brane.

\subsection{Initial theory}

We will start with definition of the background fields of the initial theory and introduce effective vector background fields and effective field strengths for a three-torus with $D_1$-brane.

\subsubsection{Background fields of initial theory}

The coordinates of the $D=3$ dimensional torus are denoted by $x^{0},x^{1},x^{2}$. In our particular example, nontrivial components of the  background are
\begin{eqnarray}\label{eq:nasapolja}
G_{\mu\nu}=\left(
\begin{array}{ccc}
1 & 0 & 0\\
0 & -1 & 0\\
0 & 0 & -1
\end{array}
\right)  \, , \qquad
B_{\mu\nu} =\left(
\begin{array}{ccc}
0 & B & 0\\
-B & 0 & 0\\
0 & 0 & 0
\end{array}
\right) \, ,
\end{eqnarray}
which produce
\begin{eqnarray}
\Pi_{\pm \mu\nu}  \equiv B_{\mu\nu}  \pm \frac{1}{2} G _{\mu\nu}
=  \frac{1}{2} \left(
\begin{array}{ccc}
\pm 1 & 2 B  & 0\\
-2 B  & \mp 1 & 0\\
0 & 0 & \mp 1
\end{array}
\right)   \, .
\end{eqnarray}

We will examine a $D_1$-brane defined with the Dirichlet boundary conditions  $x^2 (\tau, \sigma)/_{\sigma=0} = x^2 (\tau, \sigma)/_{\sigma=\pi} = const$. This means that according to our convention we will have $p=1$,  $a,b=0,1$ and $i,j=2$.  So,
we will work with Neumann background fields $A_N^0$ and  $A_N^1$ and Dirichlet background field $A_D^2$.

Such a configuration  produces $\gamma^{(0)}_2 = \kappa x^{\prime 2}$ and the action (\ref{eq:Sopen}) takes the form
\begin{eqnarray}\label{eq:Sopenpr}
& S_{open} [x] =  \kappa \int_{\Sigma} d^2\xi   \partial_{+}x^{\mu}      \Pi_{+ \mu\nu}  \partial_{-}x^{\nu}     \\
+&  2\kappa \int d\tau \Big[ \Big( A_0^N [x] {\dot x}^0 +  A_1^N [x] {\dot x}^1  + {A}_2^D [x]  x^{\prime 2} \Big)/_{\sigma=\pi}
- \Big( A_0^N [x] {\dot x}^0 +  A_1^N [x] {\dot x}^1  + {A}_2^D [x]  x^{\prime 2}    \Big)/_{\sigma=0} \Big] \,  .  \nonumber
\end{eqnarray}
Note an unusual coupling of ${A}_2^D$ with $x^{\prime 2}$.

It is easy to find  effective metric and non-commutativity parameters
\begin{eqnarray}\label{eq:efmncp}
G^E_{\mu\nu}=\left(
\begin{array}{ccc}
G_E & 0 & 0\\
 0 & -G_E & 0\\
0 & 0 & -1
\end{array}
\right)  \, , \qquad
\theta_{\mu\nu} =\left(
\begin{array}{ccc}
0 & \theta & 0\\
- \theta & 0 & 0\\
0 & 0 & 0
\end{array}
\right) \, ,
\end{eqnarray}
where
\begin{eqnarray}\label{eq:GEt}
G_E \equiv  1-4 B^2  \,  , \qquad   \theta \equiv \frac{2 B}{\kappa G_E} \, .
\end{eqnarray}
We will also need an expression for the combination of background fields,
\begin{eqnarray}\label{eq:tpm}
\theta_\pm^{\mu\nu} = \theta^{\mu\nu} \mp \frac{1}{\kappa} G^{-1 \mu \nu}_E  =
\left(
\begin{array}{ccc}
\mp \frac{1}{\kappa G_E} & \theta & 0\\
- \theta & \pm \frac{1}{\kappa G_E}   & 0\\
0 & 0 & \pm \frac{1}{\kappa}
\end{array}
\right) \, .
\end{eqnarray}

According to Eq.~(\ref{eq:Alc})  the nontrivial vector background fields are:
\begin{eqnarray}\label{eq:AlcP}
A_0^N (x)= A^0_0 - \frac{1}{2} F^{(a)} x^1 \,  , \qquad  A_1^N (x)= A^0_1 +  \frac{1}{2} F^{(a)} x^0    \,  , \qquad       {A}_2^D (x) = A^0_2 - \frac{1}{4} F^{(s)} x^2 \,   \,  ,
\end{eqnarray}
where $F^{(a)} \equiv   F_{0 1}^{(a)}= \partial_0 A^N_1 - \partial_1 A^N_0 $ and $F^{(s)} \equiv   F_{2 2}^{(s)} = - 4 \, \partial_2 A^D_2$. Consequently, the field strength of the initial theory is
\begin{eqnarray}\label{eq:fstr}
F_{\mu\nu} = F^{(a)}_{\mu\nu} + \frac{1}{2} F^{(s)}_{\mu\nu}  =
\left(
\begin{array}{ccc}
0  & F^{(a)}  & 0\\
- F^{(a)}  &  0 & 0\\
0 & 0 &  \frac{1}{2} F^{(s)}
\end{array}
\right) \, .
\end{eqnarray}
Note the unusual  appearance of symmetric field strength $F^{(s)}$.

\subsubsection{Effective vector background fields and effective field strength}

We introduce effective vector background fields, which in our  example of a three-torus with $D_1$-brane  take the forms of Eqs.~(\ref{eq:calA})  and (\ref{eq:calAeq}),
\begin{eqnarray}\label{eq:effvbf}
{\cal  A}_{\pm 0} (x)=  A^N_0 (x)  \,  , \qquad   {\cal  A}_{\pm 1} (x)=  A^N_1 (x)     \,  , \qquad  {\cal  A}_{\pm 2} (x)=  \mp   A^D_2 (x)    \,   \,  ,
\end{eqnarray}
or in components,
\begin{eqnarray}\label{eq:effvbfc}
& {\cal  A}_{0 0} (x)=  A^N_0 (x)  \,  , \qquad   {\cal  A}_{0 1} (x)=  A^N_1 (x)     \,  , \qquad   {\cal  A}_{1 0} (x)= 0 =  {\cal  A}_{1 1} (x)   \,  , \qquad \\
& {\cal  A}_{0 2} (x)= 0   \,  , \qquad   {\cal  A}_{1 2} (x)=   -   A^D_2 (x)    \,   \,  .
\end{eqnarray}
Note that the constraints on the effective fields, Eqs.~(\ref{eq:cvf}) and (\ref{eq:cvfr}), are satisfied.

The effective field strength is equivalent to the initial field strength in Eqs.~(\ref{eq:FcF0}), (\ref{eq:FcF}) and (\ref{eq:FcFs}):
\begin{eqnarray}\label{eq:efffs}
{\cal  F}_{\mu \nu} = F_{\mu \nu}   \,  , \qquad   {\cal  F}_{\mu \nu}^{(a)} =  \partial_\mu {\cal  A}_{0 \nu} (x) - \partial_\nu {\cal  A}_{0 \mu} (x)    \,  ,  \qquad  \\
 {\cal  F}_{0 0}^{(s)} = 0    \,  ,  \qquad    {\cal  F}_{1 1}^{(s)} = 0    \,  ,  \qquad   {\cal  F}_{2 2}^{(s)} =  4   \partial_2 {\cal  A}_{1 2} (x)     \,   .
\end{eqnarray}

%2 \Big( \partial_2 {\cal  A}_{1 2} (x) + \partial_2 {\cal  A}_{1 2} (x)  \Big)  =

\subsection{T-dual  theory}

Using the method described above, we will compute background fields and field strengths after  T-dualization along all coordinates. We will obtain a T-dual three-torus with $D_0$-brane.

\subsubsection{Background of T-dual theory}

According to Eq.~(\ref{eq:solV}),
in T-dual theory the vector background fields depend not only on the dual coordinate $y_\mu$ but on the expression
\begin{eqnarray}\label{eq:V}
V^{\mu}   =
\left(
\begin{array}{c}
\frac{1}{G_E} (-2 B y_1 + {\tilde y}_0)  \\
\frac{1}{G_E} (2 B y_0 - {\tilde y}_1 ) \\
- {\tilde y}_2
\end{array}
\right) \, ,
\end{eqnarray}
where ${\tilde y}_\mu$ is defined in Eq.~(\ref{eq:AtA}).  The T-dual action (\ref{eq:tdual}) takes the form
\begin{eqnarray}\label{eq:tdualpr}
& {}^\star S [y] =   \,
\frac{\kappa^{2}}{2}  \int d^{2}\xi\ \partial_{+}y_\mu \theta_{-}^{\mu\nu} \partial_{-}y_\nu  \nonumber \\
&+ 2 \kappa \int d \tau \Big[ \Big( -   {A}_2^D ( V)  {\dot y}_2  - \frac{1}{G_E}  A_0^N (V) (2B {\dot y}_1 - y^\prime_0)    - \frac{1}{G_E}  A_1^N (V) (-2B {\dot y}_0+ y^\prime_1)  ) \Big)/_{\sigma=\pi} \nonumber \\
&- \Big( -   {A}_2^D ( V)  {\dot y}_2  - \frac{1}{G_E}  A_0^N (V) (2B {\dot y}_1 - y^\prime_0)    - \frac{1}{G_E}  A_1^N (V) (-2B {\dot y}_0+ y^\prime_1)   \Big)/_{\sigma=0}       \Big] \,  ,
\end{eqnarray}
where we used Eq.~(\ref{eq:dualmom}) for $\sigma$-momenta,
\begin{eqnarray}\label{eq:tdg}
{}^\star \gamma^0_{(0)} = \frac{\kappa}{G_E} (2 B {\dot y}_1 - y^\prime_0)  \,  ,  \qquad  {}^\star \gamma^1_{(0)} = \frac{\kappa}{G_E} (- 2 B {\dot y}_0 + y^\prime_1)  \,  .
\end{eqnarray}

Consequently, according to Eq.~(\ref{eq:tdbfv})  the T-dual background fields are
\begin{eqnarray}\label{eq:nasapoljaE}
{}^\star  G^{\mu\nu}= G_E^{-1 \mu \nu} =
\left(
\begin{array}{ccc}
\frac{1}{G_E} & 0 & 0\\
0 & - \frac{1}{G_E} & 0\\
0 & 0 & -1
\end{array}
\right)  \, , \qquad
{}^\star B^{\mu\nu} = \frac {\kappa}{2} \theta^{\mu\nu} =
\frac {\kappa}{2} \left(
\begin{array}{ccc}
0 & \theta & 0\\
-\theta & 0 & 0\\
0 & 0 & 0
\end{array}
\right) \, ,
\end{eqnarray}
and
\begin{eqnarray}\label{eq:tdbf1}
{}^\star  A_D^0 (V)= \frac{1}{G_E}  A^N_0 (V)     \,  , \qquad   {}^\star A_D^1 (V)= - \frac{1}{G_E}  A^N_1 (V)    \,   \, , \qquad  {}^\star  A^2_N (V)= -  A^D_2 (V) \,     .
\end{eqnarray}
Note that now we have one T-dual  Neumann and two T-dual Dirichlet vector fields. It means that the T-dual three-torus has a $D_0$-brane
defined with the Dirichlet boundary conditions   $y_0 (\tau, \sigma)/_{\sigma=0} = y_0 (\tau, \sigma)/_{\sigma=\pi} = const$ and
$y_1 (\tau, \sigma)/_{\sigma=0} =  y_1 (\tau, \sigma)/_{\sigma=\pi} = const$, as well as   ${\tilde y}_0 (\tau, \sigma)/_{\sigma=0} = {\tilde y}_0 (\tau, \sigma)/_{\sigma=\pi} = const$ and
${\tilde y}_1 (\tau, \sigma)/_{\sigma=0} = {\tilde y}_1 (\tau, \sigma)/_{\sigma=\pi} = const$.

The T-dual effective vector background fields in term of initial ones are ${}^\star {\cal  A}_\pm^\mu = \kappa \, \theta^{\mu\nu}_\pm {\cal  A}_{\pm \nu}$ which, with the help of Eq.~(\ref{eq:effvbf}),
is equivalent to Eq.~(\ref{eq:calAGd}).
So, in  the case of the present example we have:
\begin{eqnarray}\label{eq:efca}
& {}^\star {\cal  A}_\pm^0  (V)=  \mp \frac{1}{G_E} {\cal  A}_{\pm 0} (V) + \kappa \, \theta  {\cal  A}_{\pm 1} (V) = \mp \frac{1}{G_E} A^N_0 (V)  + \kappa \, \theta   A^N_1 (V)   \,  , \\
& {}^\star {\cal  A}_\pm^1  (V)=  - \kappa \, \theta  {\cal  A}_{\pm 0} (V)   \pm \frac{1}{G_E} {\cal  A}_{\pm 1} (V)  = - \kappa \, \theta   A^N_0 (V) \pm  \frac{1}{G_E} A^N_1 (V)    \,  ,   \\
& {}^\star {\cal  A}_\pm^2  (V)=  \pm {\cal  A}_{\pm 2} (V) =  -   A^D_2 (V)    \,   \,  .
\end{eqnarray}

Rewriting this in components, or according to Eq.~(\ref{eq:calAcd}), we obtain:
\begin{eqnarray}\label{eq:efcac}
& {}^\star {\cal  A}_0^0  (V) =    \kappa \, \theta   A^N_1 (V)   \,  ,  \qquad    {}^\star {\cal  A}_0^1  (V) =  - \kappa \, \theta   A^N_0 (V) \,  ,  \qquad    {}^\star {\cal  A}_0^2  (V) =  - A^D_2 \,  ,  \\
&  {}^\star {\cal  A}_1^0  (V) = - \frac{1}{G_E}  A^N_0 (V) \,  ,  \qquad      {}^\star {\cal  A}_1^1  (V) =  \frac{1}{G_E}  A^N_1 (V) \,  ,  \qquad     {}^\star {\cal  A}_1^2  (V) = 0                \,  ,
\end{eqnarray}
where $G_E$ and $\theta$ are defined in Eq.~(\ref{eq:GEt}).  Note that the constraints of Eq.~(\ref{eq:cvfd}) are satisfied.

\subsubsection{T-dual transformation laws}

The T-dual transformation laws from a three-torus with $D_1$-brane  to dual three-torus with $D_0$-brane,  in accordance with  Eq.~(\ref{eq:tdtr}),  take the form
\begin{eqnarray}\label{eq:trl}
&  \partial_\pm x^0 \cong \pm \frac{1}{G_E} ( \partial_\pm y_0 \mp 4 {\cal A}_{\pm 0}) - \kappa \theta ( \partial_\pm y_1 \mp 4 {\cal A}_{\pm 1})    \,  , \nonumber \\
& \partial_\pm x^1 \cong  \kappa \theta ( \partial_\pm y_0 \mp 4 {\cal A}_{\pm 0})  \mp \frac{1}{G_E} ( \partial_\pm y_1 \mp 4 {\cal A}_{\pm 1})     \,  ,  \nonumber \\
&   \partial_\pm x^2 \cong  \mp ( \partial_\pm y_2 \mp 4 {\cal A}_{\pm 2})   \,   \,  ,
\end{eqnarray}
while the inverse, from Eq.~(\ref{eq:itdtr}), is
\begin{eqnarray}\label{eq:trli}
&  \partial_\pm y_0 \cong \pm  \partial_\pm x^0  - 2 B \partial_\pm x^1  \pm   4 {\cal A}_{\pm 0}          \,  , \nonumber  \\
&  \partial_\pm y_1 \cong 2 B  \partial_\pm x^0  \mp \partial_\pm x^1  \pm   4 {\cal A}_{\pm 1}        \,  ,  \nonumber \\
&  \partial_\pm y_2 \cong \mp  \partial_\pm x^2   \pm   4 {\cal A}_{\pm 2}       \,   \,  .
\end{eqnarray}

Note that the expression for $V^\mu$  (\ref{eq:V}) is a solution of the finite part  of Eq.~(\ref{eq:trl}), for  ${\cal A}_{\pm 0} = {\cal A}_{\pm 1} =0$.

\subsubsection{T-dual field strength}

In Eq.~(\ref{eq:fstr})  we introduced the field strength of the vector background fields on the string end-points for initial theory of a
three-torus with $D_1$-brane. Now, we are going to express the field strength of its T-dual three-torus with $D_0$-brane obtained  after dualization   along all coordinates.

In the T-dual Dirichlet sector, with the help of   Eq.~(\ref{eq:tpm}),  we have
\begin{eqnarray}\label{eq:ktft}
\kappa \, \theta^{\mu\rho}_\pm \, {\cal F}^{(a)}_{\rho \sigma} \, \kappa \, \theta^{\sigma \nu}_\pm  =
{{\cal F}^{(a)} \over G_E^2}
\left(
\begin{array}{ccc}
\pm 4 B & -(1+ 4 B^2)  & 0\\
1+ 4 B^2  &  \mp 4 B & 0\\
0 & 0 &  0
\end{array}
\right) \, ,
\end{eqnarray}
where ${\cal F}^{(a)} \equiv {\cal F}^{(a)}_{01} = F^{(a)}_{01} \equiv F^{(a)}$. So, according to Eqs.~(\ref{eq:dFcFDic}) and (\ref{eq:dFcFDisc}) we find
\begin{eqnarray}\label{eq:dFa}
{}^\star {\cal F}_{(a)}^{a b}   =
- {{\cal F}^{(a)} \over G_E^2}
\left(
\begin{array}{ccc}
0& -(1+ 4 B^2)  & 0\\
1+ 4 B^2  &  0 & 0\\
0 & 0 &  0
\end{array}
\right) \, ,
\end{eqnarray}
and
\begin{eqnarray}\label{eq:dFs}
{}^\star {\cal F}_{(s)}^{a b}   =
2 {{\cal F}^{(a)} \over G_E^2}
\left(
\begin{array}{ccc}
4 B & 0  & 0\\
0  &  - 4 B & 0\\
0 & 0 &  0
\end{array}
\right) \, .
\end{eqnarray}

The only nontrivial term in the Neumann sector is the second term in Eq.~(\ref{eq:dFcFDi0s}),  ${}^\star {\cal F}_{(s)}^{2 2} = - {\cal F}^{(s)}_{2 2} = - F^{(s)}_{2 2} \equiv - F^{(s)}$.
Consequently, the complete field strength of the T-dual three-torus with $D_0$-brane is
\begin{eqnarray}\label{eq:cf}
{}^\star {\cal F}^{\mu \nu}   = {}^\star {\cal F}^{\mu \nu}_{(a)} + \frac{1}{2}  {}^\star {\cal F}^{\mu \nu}_{(s)} =
\left(
\begin{array}{ccc}
{4 {\cal F}^{(a)} B \over G_E^2} &  {{\cal F}^{(a)} (1+ 4 B^2) \over G_E^2}     & 0\\
- {{\cal F}^{(a)} (1+ 4 B^2) \over G_E^2} &  - {4 {\cal F}^{(a)} B \over G_E^2} & 0\\
0 & 0 &  - {\cal F}^{(s)}
\end{array}
\right) \, ,
\end{eqnarray}
where
${\cal F}^{(a)} \equiv {\cal F}^{(a)}_{0 1} = \partial_0 A^N_1 - \partial_1 A^N_0$ and
${\cal F}^{(s)} \equiv {\cal F}^{(s)}_{2 2} = - 4 \partial_2 A^D_2$. In terms of T-dual fields we have
${\cal F}^{(a)}  = - G_E \, (\partial_0 {}^\star  A_D^1 + \partial_1 {}^\star A_D^0)$ and
${\cal F}^{(s)}  =  4 \, \partial_2 {}^\star A_N^2$.

Since in our particular case ${\cal F}^{(a)} = F^{(a)}$ and ${\cal F}^{(s)} = F^{(s)}$, we can write
\begin{eqnarray}\label{eq:cff}
{}^\star {\cal F}^{\mu \nu}   = {}^\star {\cal F}^{\mu \nu}_{(a)} + \frac{1}{2}  {}^\star {\cal F}^{\mu \nu}_{(s)} =
\left(
\begin{array}{ccc}
{4 {F}^{(a)} B \over G_E^2} &  {F^{(a)} (1+ 4 B^2) \over G_E^2}     & 0\\
- {F^{(a)} (1+ 4 B^2) \over G_E^2} &  - {4 {F}^{(a)} B \over G_E^2} & 0\\
0 & 0 &  - {F}^{(s)}
\end{array}
\right) \, ,
\end{eqnarray}
where $F^{(a)}$ and $F^{(s)}$ have been introduced after Eq.~(\ref{eq:AlcP}). Again, besides  antisymmetric field strength we have a non-trivial symmetric part of field strength ${F}^{(s)}$.

\subsubsection{Non-geometric three-torus with $D_0$-brane}

The basic relation (\ref{eq:cfsF}) of the field strength of non-geometric theories can be expressed in the form
\begin{eqnarray}\label{eq:fsng}
{}^\star {\cal F}^{\mu \nu}   =
{\hat \partial}^\mu_+ {}^\star {\cal A}^\nu_+ -  {\hat \partial}^\nu_- {}^\star {\cal A}^\mu_-     \, ,
\end{eqnarray}
where ${\hat \partial}^\mu_\pm = \partial^\mu_y \pm \partial^\mu_{\tilde y}$ and ${}^\star {\cal A}^\mu_\pm = {}^\star {\cal A}^\mu_0  \pm {}^\star {\cal A}^\mu_1$.

In the case of a three-torus it becomes the expression
\begin{eqnarray}\label{eq:fstt}
{}^\star {\cal F}^{\mu \nu}  =
\left(
\begin{array}{ccc}
 {\hat \partial}^0_+ {}^\star {\cal A}^0_+ - {\hat  \partial}^0_- {}^\star {\cal A}^0_-    &   {\hat \partial}^0_+ {}^\star {\cal A}^1_+ - {\hat  \partial}^0_- {}^\star {\cal A}^1_-
 &  {\hat \partial}^0_+ {}^\star {\cal A}^2_+ - {\hat  \partial}^0_- {}^\star {\cal A}^2_- \\
 {\hat \partial}^1_+ {}^\star {\cal A}^0_+ - {\hat  \partial}^1_- {}^\star {\cal A}^0_-    &   {\hat \partial}^1_+ {}^\star {\cal A}^1_+ - {\hat  \partial}^1_- {}^\star {\cal A}^1_-
 &  {\hat \partial}^1_+ {}^\star {\cal A}^2_+ - {\hat  \partial}^1_- {}^\star {\cal A}^2_-           \\
{\hat \partial}^2_+ {}^\star {\cal A}^0_+ - {\hat  \partial}^2_- {}^\star {\cal A}^0_-    &   {\hat \partial}^2_+ {}^\star {\cal A}^1_+ - {\hat  \partial}^2_- {}^\star {\cal A}^1_-
 &  {\hat \partial}^2_+ {}^\star {\cal A}^2_+ - {\hat  \partial}^2_- {}^\star {\cal A}^2_-
\end{array}
\right) \, .
\end{eqnarray}

The term with effective background fields of  T-dual action  (\ref{eq:sSay}) takes the form
\begin{eqnarray}\label{eq:sSayE}
{}^\star S_A [y] =
2\kappa  \int d\tau \Big[ \Big( ^\star {\cal A}^\mu_0 [V] \,  {\dot y}_\mu - ^\star {\cal A}^\mu_1 [V] \,  {y}_\mu^\prime   \Big)/_{\sigma=\pi}  -
\Big( ^\star {\cal A}^\mu_0 [V] \,  {\dot y}_\mu - ^\star {\cal A}^\mu_1 [V] \,  {y}_\mu^\prime   \Big)/_{\sigma=0}     \Big)    \Big]  \,  \,  .
\end{eqnarray}
Note that $^\star {\cal A}^\mu_1 [V]$ is multiplied by ${y}_\mu^\prime $, not by ${\dot y}_\mu$.

Let us stress that only the first term in Eq.~(\ref{eq:fsng}) ${}^\star {\cal F}^{\mu \nu}_{standard}   =
{\partial}^\mu_y {}^\star {\cal A}^\nu_0 -  {\partial}^\nu_y {}^\star {\cal A}^\mu_0 $
is the standard one. The other three terms are new and unusual. One is also antisymmetric, but with derivation with respect to  ${\tilde y}_\mu$, while the
other two are symmetric.

\subsection{Genuinely non-geometric three-torus}

If we preserve the relation between effective background fields we can go back to the initial geometric  theory.
As explained at the beginning of Section~6, if we want to introduce the new kind of theory we can use the obtained structure from Eqs.(\ref{eq:fsng})-(\ref{eq:sSayE}) and suppose that: firstly, the effective background fields and corresponding  field strengths are independent; and secondly,
that background fields depend on the solutions of T-duality transformations. In that case we will lose the possibility to go back to the initial theory with inverse T-dualization. So, all
our theories will be genuinely non-geometric.

With the constraints for effective vector fields the Neumann part of ${\cal A}_{1 \mu}$ is zero,
\begin{eqnarray}\label{eq:Npec}
{\cal A}_{1 \mu}  =
\left(
\begin{array}{c}
0   \\
 0    \\
- A^D_2
\end{array}
\right) \, ,
\end{eqnarray}
but without constraints we have the non-trivial expression
\begin{eqnarray}\label{eq:Npe}
{\cal A}_{1 \mu}  =
\left(
\begin{array}{c}
2 B \, ^\star {\cal A}^1_1 - \, ^\star {\cal A}^0_0    \\
- 2 B \, ^\star {\cal A}^0_1 + \, ^\star {\cal A}^1_0      \\
 ^\star {\cal A}^2_0
\end{array}
\right) \,  .
\end{eqnarray}

The initial vector fields  ${\cal A}_{\pm \mu}$  depend on the solution of the finite part of inverse  T-dual transformation laws (\ref{eq:trli}),
\begin{eqnarray}\label{eq:Vx}
V_{\mu}  (x)  =
\left(
\begin{array}{c}
-2 B x^1 + {\tilde x}^0  \\
2 B x^0 - {\tilde x}^1  \\
- {\tilde x}^2
\end{array}
\right) \, .
\end{eqnarray}
The T-dual ones  $^\star {\cal A}^\mu_\pm$, depend  on $V^\mu (y)$ defined in Eq.~(\ref{eq:V}), which is the solution of the finite part of the T-dual transformation laws (\ref{eq:trl}).

The local gauge symmetry is defined with Eq.~(\ref{eq:dnggtc}), while the matter fields can be introduced  with Eqs.~(\ref{eq:Lag}) and (\ref{eq:Lagi}).

%%%%%%%%%%%%%%%%%%%%%%%%%%%%%%%%%%%%%%%%%%%%%%%%%%%%%%%%%%%%%%%%%%%%%%%%%%%%%%%%%%%%%
\cleq

\section{ Conclusions}

In the present article, using T-duality of the vector fields, we are able to introduce a new definition for a geometrical feature (the field strength) in  non-geometric (T-dual) theories.

We started with T-duality of the vector gauge fields. In string theory the  gauge fields appear at the boundaries of the open string. Their role is to  enable complete local gauge symmetries.
In fact, there are two important symmetries of the closed  string theory: local gauge symmetry of the Kalb-Ramond field and general coordinate transformations.
In Section 3.2 we showed that the symmetry T-dual to local gauge symmetry  includes transformations of the background fields  but does not include transformations of the string coordinates. Both symmetries fail at the open string end-points.
The function of gauge fields is to restore these symmetries  at the end-points. So, they are defined only on the open string boundary and not on the whole world-sheet.
The corresponding term in the action is a line integral over the world-sheet boundary.

To each of the above symmetries  of the  string theory  there corresponds an appropriate vector gauge field.  As a consequence of the boundary conditions only parts of these gauge fields survive.
From a gauge field corresponding to local gauge symmetry of the Kalb-Ramond field,  the components along coordinates with Neumann boundary conditions survive.
From  a gauge field corresponding to restricted general coordinate transformations,  the components along coordinates with Dirichlet  boundary conditions survive.
So, we obtained one complete vector field $\{A^N_a , A^D_i \}, \,\, \mu= (a,i)$. The action which describes  field $A^N_a$ is a standard one (see for example Ref.~\cite{Zw}), while  introduction of the action for field $A^D_i$ is the contribution of the
present article.

There are several  important results in the present article. First, we  added a new term  ${A}_i^D [x] G^{-1 i j}  \gamma^{(0)}_j (x)/_{\partial \Sigma}$
in the action  (\ref{eq:Sopen})  which corresponds to the Dirichlet boundary conditions and which compensates
not-implemented  general coordinate transformations at string end-points. We considered the case when the vector gauge field is linear in coordinates, so that it satisfies the open string space-time equations of motion.

Second, we perform T-duality along all the coordinates. We used a new approach  for T-dualization in the absence of global symmetry \cite{DS2}. We showed that such T-dualization exchanges:
1) Neumann with Dirichlet boundary conditions; 2) initial Dirichlet vector fields $A^D_i (x)$ with T-dual Neumann vector fields ${}^\star A_N^i (V)$
(also initial Neumann vector fields $ A^N_a (x)$ with  T-dual Dirichlet vector fields ${}^\star A_D^a (V)$);
 and 3) local gauge transformations with general coordinate transformations. Note that in initial  theory the gauge fields depend on $x^\mu$ while in  T-dual non-geometric theory  they  depend on the non-local variable $V^\mu$.
 This is the cause of many interesting consequences.

Third,  we introduced  field strength for T-dual theories.
The final expression is  in accordance with the result obtained in another way, with direct T-dualization of the action with field strength.
Using the fact that T-duality transformation turns geometric to  non-geometric theories we can express effective T-dual field strength ${}^\star {\cal F}^{\mu \nu}$ as a derivation of effective
 gauge fields  ${}^\star {\cal A}^\mu_\alpha (V)$, see Eq.~(\ref{eq:cfsF}).
Because the arguments of non-geometric theories depend on $V^\mu = - \kappa \, \theta^{\mu \nu} y_\nu + G^{-1 \mu \nu}_E \, {\tilde y}_\nu$, the corresponding field strength contains derivatives with respect to both $y_\mu$ and
${\tilde y}_\mu$. We also find  that field strength (\ref{eq:cfsF}) is invariant under  gauge transformations of non-geometric theories (\ref{eq:nggtc}).

Fourth, when we omit the relation between effective background fields ${\cal A}^\mu_\alpha (V)$,  proclaiming  them independent,  and introduce new arguments of background fields as a solution of T-dual transformation laws,
we will not be able to go back to the initial  geometric theory.  So, all our theories in any duality frame become truly  non-geometric.

In another paper  \cite{BS1}  we reproduce the results of the present article in the double space introduced in Ref.~\cite{BS}.
Let us stress that there is an essential difference between our approach and that of double field theories  \cite{HZ,HHZ}. In double field theories  there are two coordinates, the initial $x^\mu$ and its double, denoted as $\tilde x_\mu$. The variable  $\tilde x_\mu$ corresponds to our $y_\mu$ but we have an additional dual coordinate $\tilde y_\mu$ defined in first relation of Eq.~(\ref{eq:AtA}). It plays an essential role in the definition of field strength for non-geometric theories.

It will be interesting to establish a relation between our formulation  and the recent work of other authors  on double field theory.
The fact that  double field theory does not depend on ${\tilde y_\mu}$ suggests that in order to find  a relation between these theories we should eliminate the variable ${\tilde y_\mu}$,  expressing it in terms of $y_\mu$.
For example, it is  possible to introduce a Lagrange multiplier $\lambda^\mu$ and add the term $\lambda^\mu ({\dot y_\mu} - {{\tilde y_\mu}}^\prime)$ to the Lagrangian
 in order to introduce the relation between these variables as a constraint. Then ${\tilde y_\mu}$ becomes an independent variable, but we pay the price of introducing a
new variable $\lambda^\mu$.

There may be a new, more general theory such that both double field theory and our theory formulated in  double space are some particular cases of that general theory.

Consequently, in the present paper we extended features of non-geometric theories to the case of an open string. In the formulation with gauge fields, similar to the closed string case, the  non-geometry can be noticed in
arguments of T-dual gauge fields  (non-local expressions of $V^a$). Non-standard couplings of Dirichlet field, not only with  ${\dot x}^\mu$ but also with  $x^{\prime \mu}$, should also  be mentioned.
In the formulation with field strength, the T-dual  field strength is derivative of vector gauge fields with respect to both the T-dual variable $y_\mu$ and its double ${\tilde y}_\mu$.
It  depends not only on the antisymmetric but also on the symmetric  part. This can be seen from the T-dual expression for field strength,
Eq.~(\ref{eq:cfsF}),  which is the  main contribution of this paper.
All these features  are completely new   and in any case they  require  further investigations.

%%%%%%%%%%%%%%%%%%%%%%%%%%%%%%%%%%%%%%%%%%%%%%%%%


\begin{thebibliography}{99}


\bibitem{HMW} S. Hellerman, J. McGreevy and B. Williams, {\it JHEP}, {\bf 01}: 024 (2004)

\bibitem{DH} A. Dabholkar and C. Hull, {\it JHEP}, {\bf 09}:  054 (2003)

\bibitem{STW} J. Shelton, W. Taylor and B. Wecht, {\it JHEP}, {\bf 10}:  085 (2005)

\bibitem{H} Hull, {\it JHEP}, {\bf 065}:  0510  (2005)

\bibitem{DS1} Lj. Davidovi\'c and B. Sazdovi\'c, {\it  Eur. Phys. J. C },  {\bf 74}:2683 (2014)

\bibitem{DS2} Lj. Davidovi\'c and B. Sazdovi\'c, {\it JHEP}, {\bf 11}:  119 (2015)

\bibitem{DNS} Lj. Davidovi\'c, B. Nikoli\'{c} and B. Sazdovi\'c, {\it  Eur. Phys. J. C},   {\bf 74}:2734 (2014)

\bibitem{DNS2} Lj. Davidovi\'c, B. Nikoli\'{c} and B. Sazdovi\'c, {\it  Eur. Phys. J. C},   {\bf 7}5:576 (2015)

\bibitem{SW}  Seiber and Witten, {\it JHEP},  {\bf 032}:  9909  (1999)

\bibitem{SA1} R. J. Szabo,   {\it Int. J. Mod. Phys. A}, {\bf 19}: 1837 (2004)


\bibitem{DL}  D. Lust, {\it JHEP}, {\bf 12}:  084 (2010)

\bibitem{BDLPR} R. Blumenhagen, A. Deser, D. L\"ust, E. Plauschinn and F. Rennecke, {\it J. Phys. A}, {\bf 44}:  385401 (2011)

\bibitem{SA2}  R. J. Szabo,   {\it Class. Quant. Grav.}, {\bf 23}:  R199  (2006)


\bibitem{Zw} B. Zwiebach, {\it A First Course in String Theory}, Second edition (Cambridge University Press, 2002), p. 673

\bibitem{Le} R. G. Leigh, {\it Mod. Phys. Lett. A }, {\bf 4}:  2767 (1989)

\bibitem{Pol} J. Polchinski, {\it String theory, Volume I}, First edition (Cambridge University Press, 1998), p. 402


\bibitem{B} T. Buscher, {\it Phys. Lett.B }, {\bf 194}: 51 (1987); {\bf 201}  466 (1988)


\bibitem{EO} M. Evans and B. A Ovrut, {\it Phys.Rev. D}, {\bf 39}:  3016 (1989)

\bibitem{EO1} M. Evans and B. A Ovrut, {\it Phys.Rev. D}, {\bf 41}:  3149 (1990)


\bibitem{DS3} Lj. Davidovi\'c and B. Sazdovi\'c, in preparation.



\bibitem{BHM}  P. Bouwknegt, K. Hannabuss, and V.  Mathai,  {\it Commun.Math.Phys.}, {\bf 264}:   41 (2006)


\bibitem{BMRS}  J. Brodzki, V. Mathai, J. Rosenberg, and R. J. Szabo, {\it Commun.Math.Phys.}, {\bf 277}: 643 (2008)



\bibitem{BP} R. Blumenhagen and  E. Plauschinn, {\it J. Phys. A}, {\bf 44}:  015401 (2011).

\bibitem{S} K. Becker, M. Becker and J. Schwarz, {\it String Theory and M-Theory: A Modern Introduction}, First edition (Cambridge University Press,  2007), p. 739


\bibitem{DO} H. Dorn and H.-J. Otto, {\it Phys. Lett. B}, {\bf 381}:  81 (1996)

\bibitem{ABB} E. Alvarez, J.L.F. Barbon and J. Borlaf, {\it Nucl. Phys. B}, {\bf 479}:  218 (1996)

\bibitem{CJL} A. Chatzistavrakidis, L. Jonke and O. Leehtenfeld, {\it JHEP}, {\bf 11}:  182 (2015)

\bibitem{BS1} B. Sazdovi\'c, {\it Eur. Phys. J. C},  {\bf 77}: 634  (2017)


\bibitem{BS} B. Sazdovi\'c, {\it JHEP}, {\bf 08}:  055 (2015)

\bibitem{HZ} C. Hull and B. Zwiebach,  {\it JHEP}, {\bf 09}:  099 (2009); {\it JHEP}, {\bf 09}:  090 (2009)

\bibitem{HHZ} O. Hohm, C. Hull and B. Zwiebach,  {\it JHEP}, {\bf 08}:  008 (2010)



\end{thebibliography}
\end{document}